\documentclass[aps,pra,showpacs,twocolumn,superscriptaddress]{revtex4-1}
\usepackage{amsmath}
\usepackage{amssymb}
\usepackage{graphicx}
\usepackage{subfigure}
\usepackage{epstopdf}
\usepackage{times,txfonts}
\usepackage{easybmat}
\usepackage{multirow}
\usepackage{listings}
\usepackage{diagbox}
\usepackage{mathtools}

\usepackage[bookmarks=false]{hyperref}
\hypersetup{colorlinks=true,citecolor=blue,linkcolor=blue,urlcolor=blue,pdfstartview=FitH,bookmarksopen=true}
\usepackage{color,soul}

\begin{document}
\preprint{APS/123-QED}

\title{Direct Fidelity Estimation of Quantum States Using Machine Learning}
\author{Xiaoqian Zhang}
\altaffiliation{These authors contributed equally}
\author{Maolin Luo}
\altaffiliation{These authors contributed equally}
\affiliation{School of Physics and State Key Laboratory of Optoelectronic Materials and Technologies, Sun Yat-sen University, Guangzhou 510000, China}
\author{Zhaodi Wen}
\affiliation{College of Information Science and Technology, College of Cyber Security, Jinan University, Guangzhou 510632, China}
\author{Qin Feng}
\author{Shengshi Pang}
\affiliation{School of Physics and State Key Laboratory of Optoelectronic Materials and Technologies, Sun Yat-sen University, Guangzhou 510000, China}
\author{Weiqi Luo}
\affiliation{College of Information Science and Technology, College of Cyber Security, Jinan University, Guangzhou 510632, China}
\author{Xiaoqi Zhou}
\email{zhouxq8@mail.sysu.edu.cn}
\affiliation{School of Physics and State Key Laboratory of Optoelectronic Materials and Technologies, Sun Yat-sen University, Guangzhou 510000, China}

\date{\today}

\begin{abstract}
  In almost all quantum applications, one of the key steps is to verify that the fidelity of the prepared quantum state meets expectations. In this Letter, we propose a new approach solving this problem using machine-learning techniques. Compared to other fidelity estimation methods, our method is applicable to arbitrary quantum states, the number of required measurement settings is small, and this number does not increase with the size of the system. For example, for a general five-qubit quantum state, only four measurement settings are required to predict its fidelity with $\pm1\%$ precision in a nonadversarial scenario. This machine-learning-based approach for estimating quantum state fidelity has the potential to be widely used in the field of quantum information.
\end{abstract}

\maketitle

In the field of quantum information, almost all quantum applications require the generation and manipulation of quantum states. However, due to the imperfections of equipment and operation, the prepared quantum state is always different from the ideal state. Therefore, it is a key step to evaluate the deviation of the prepared state from the ideal one in the quantum applications. Quantum state tomography (QST) \cite{1Chantasri,3Wieczorek,4Cramer,5Renes,6Gross,7Steven,8Smith,9Kalev,10Riofr,11Kyrillidis,12Shang,13Silva,14Oh,15Siddhu,16Ma,17Martnez,18Sosa} is the standard method for reconstructing a quantum state to obtain its density matrix, which can be used to calculate the fidelity of the quantum state with respect to the ideal one. In recent years, researchers have proposed compressed sensing methods \cite{6Gross,7Steven,8Smith,9Kalev,10Riofr,11Kyrillidis} to improve the efficiency of QST for the pure quantum states.
Despite the fact that compressed sensing greatly reduces the measurement resources, the measurement settings for QST still grow exponentially with the size of the system.

However, to evaluate the fidelity of a quantum state, full reconstruction of its density matrix is not needed. Recently, schemes \cite{19Lu,20Tokunaga,21Steven,54da,22Zhu,23Cerezo,24Somma,25Wang,53Mahler,55Li,56Yu,57Zhu,58Liu,59Zhu,60Wang,61Zhu,62Li,63Sam,64Zhang,65Jiang} for directly estimating the fidelity of quantum states, including the quantum state verification (QSV) method \cite{53Mahler,55Li,56Yu,57Zhu,58Liu,59Zhu,60Wang,61Zhu,62Li,63Sam,64Zhang,65Jiang} and the direct fidelity estimation (DFE) method \cite{21Steven}, have been proposed. The QSV method can determine whether a quantum state is the target state with few measurement resources, but this method is only applicable to special quantum states, such as the stabilizer states or the $W$ states, and is not applicable to general quantum states. Compared with the QSV method, the DFE method \cite{21Steven} is applicable to general quantum pure states but requires more measurement settings. In most practical experiments, the number of measurement settings has a significant impact on the total measurement time (changing measurement setting is time consuming). Both the QSV and the DFE methods assume that the measured quantum state may be prepared or manipulated by an adversary, which is valid for the case of quantum networks. For most local experiments in which the quantum devices are trusted, the imperfections of the quantum state are caused by noise and device defects, not by the adversary. As a result, our aim is to devise a direct fidelity estimation protocol for this scenario, further reducing the number of measurement settings required.

In this Letter, we use machine-learning methods \cite{27Yang,28Ma,29Lu,30Gao,31Deng,35Ren,36Xin,39Ling,40Miszczak,41Ahmed,42Ahmed} to tackle this problem. So far, machine-learning methods have been used for classification problems \cite{27Yang,28Ma,29Lu,30Gao,31Deng,35Ren} in the field of quantum information to detect the nonlocality \cite{27Yang}, steerability \cite{29Lu} and entanglement \cite{28Ma} of quantum states. In these previous works, the classification of quantum states can be performed with high accuracy using fewer measurement settings by using artificial neural networks to learn the potential information between the internal structures of the quantum state space. In this Letter, we transform the quantum state fidelity estimation problem into a classification problem, by dividing the quantum state space into different subspaces according to the value of fidelity, and then using a neural network to predict which subspace the quantum state is in to obtain an estimate of the quantum state fidelity. Compared with previous methods for direct estimation of fidelity, this method not only works for arbitrary quantum states, but also greatly reduces the number of measurement settings required.

First, Let us review how to represent the fidelity of a quantum state using the Pauli operators. The fidelity \cite{53Tacchino} of an arbitrary quantum state $\rho$ with respect to the desired pure state $\rho_0$ can be written as
\begin{eqnarray}
\displaystyle F(\rho_0,\rho)=tr\sqrt{\rho^{1/2}\rho_0\rho^{1/2}}=\sqrt{tr(\rho\rho_0)},
\end{eqnarray}
where
\begin{eqnarray}
\rho_0=\frac{1}{2^n}\sum\limits_{j=0}^{4^n-1}a_j W_j,\quad \rho=\frac{1}{2^n}\sum\limits_{j=0}^{4^n-1}\beta_jW_j,
\end{eqnarray}
in which
\[
\sum\limits_{j=0}^{4^n-1}a_j^2/2^n=1,\quad\quad \sum\limits_{j=0}^{4^n-1}\beta_j^2/2^n\leqslant1.
\]

\vspace{-5mm}

\noindent Here $W_j$ represents Pauli operators which are $n$-fold tensor products of $I, X, Y$, and $Z$. The fidelity in Eq.(1) can be expanded in terms of the Pauli operators' expectation values $a_j$ and $\beta_j$
\begin{eqnarray}
F(\rho_0,\rho)=\sqrt{\frac{1}{2^n}\sum\limits_{j=0}^{4^n-1}\beta_j a_j}.
\end{eqnarray}

\begin{figure}[!h]
  \flushleft
  \centering
  \includegraphics[width=0.35\textwidth]{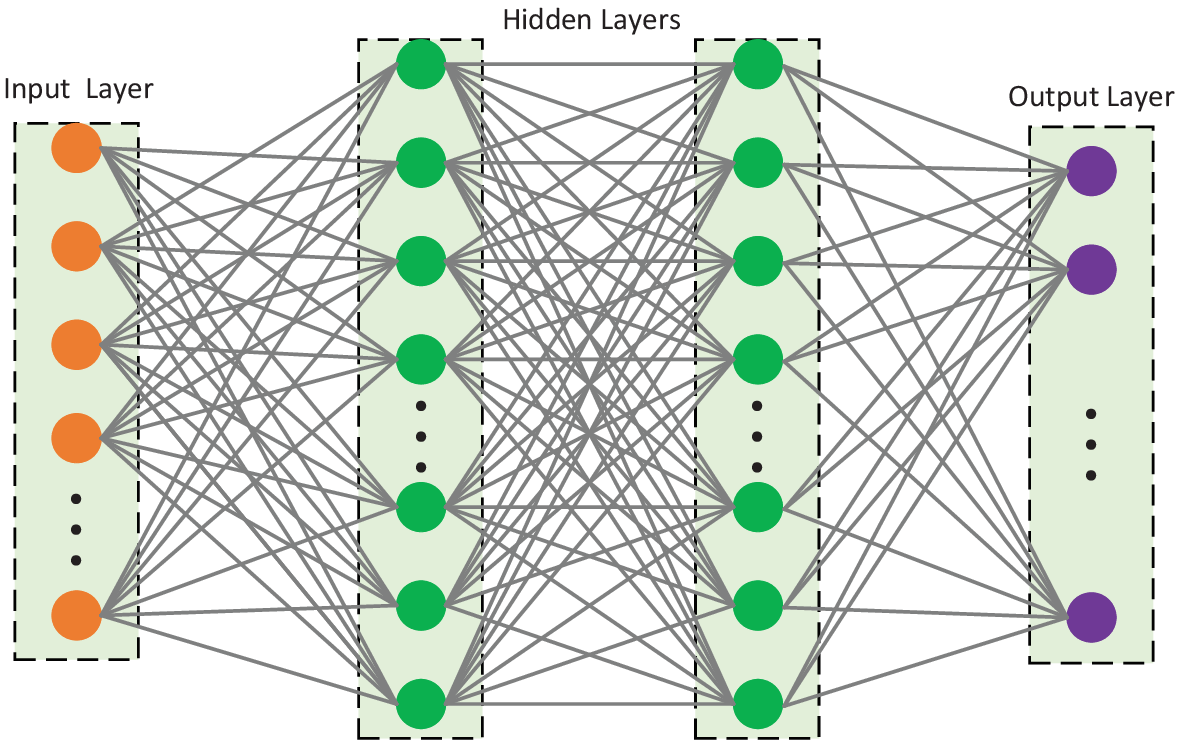}
  \caption{The artificial neural network for quantum state fidelity evaluation. The input layer neurons are loaded with the measurements of the Pauli operators, the output layer neurons correspond to different fidelity intervals, and the input and output layers are fully connected by several hidden layers. After hundreds of training sessions, a neural network model that can evaluate the fidelity of quantum states is obtained.}\label{A1}
\end{figure}

Now, we present the neural network model used for fidelity estimation. Here we choose $k$ measurement settings to measure the $n$-qubit quantum state (See Supplementary Material \cite{70SUP} Sec. III for the selection of the measurement settings). Taking a three-qubit quantum state as an example, assuming the measurement setting is XYZ, for each qubit there are two possible measurement results of +1 and -1, there will be eight possible measurement outcomes for the three-qubit state. Using these eight outcomes, it is possible to calculate the expected value of not only the Pauli operator $XYZ$, but also the expected values of the six nontrivial Pauli operators $XYI$, $XIZ$, $XII$, $IYZ$, $IYI$, and $IIZ$. For each of these $k$ measurement settings of the $n$-qubit quantum state, $2^n$ possible outcomes are obtainable. Using these $2^n$ outcomes, the expected values of the $2^n-1$ nontrivial Pauli operators can be calculated. $M$ of these $2^n-1$ expected values will be selected as neuron inputs, and thus, the input layer has a total of $k\times M$ neurons, where $M\leqslant2^n-1$.
Here we consider the case $M = 2^n-1$. We note that one can choose $M=poly(n)$ (See Supplementary Material \cite{70SUP} Sec. VIII).

Figure 1 shows the structure of the neural network we used here. The input layer neurons are fully connected to the hidden layer, i.e., each neuron of the input layer is connected to each neuron of the hidden layer. The hidden layer is also fully connected to the output layer. The output layer has a total of 122 neurons corresponding to different fidelity intervals of the quantum states. After several hundred rounds of training, the prediction accuracy of the neural network saturates, resulting in a neural network model that can predict the fidelity of the quantum states with high confidence (See Supplementary Material \cite{70SUP} Sec. VI).

In the following we describe how to generate the neural network training database for an arbitrary $n$-qubit quantum pure state. First, we generate a database for the pure state $|0\rangle^{\otimes n}\langle0|$. We divided the fidelity of the quantum state with respect to $|0\rangle^{\otimes n}\langle0|$ into 122 fidelity intervals \cite{68three}, and generated 20000 quantum states satisfying randomness and uniformity in each interval (see Supplementary Material \cite{70SUP} Sec. II), of which 16000 were used for the training of the neural network and 4000 were used for the validation of the neural network. These 2\,440\,000 quantum states then constitute our original database for the $n$-qubit quantum state fidelity estimation. For an arbitrary $n$-qubit quantum pure state $\rho_0$, a unitary matrix $U$ can be found that satisfies $\rho_0=U|0\rangle^{\otimes n}\langle0|U^\dagger$. Next, we act $U$ on each quantum state in the original database, thus obtaining a new database for $\rho_0$. Because of the invariance of the inner product of the unitary transformation, the relation of the new database with respect to $\rho_0$ is identical to that of the original database with respect to $|0\rangle^{\otimes n}\langle0|$.

Taking a general five-qubit quantum state $|\psi_0\rangle$ as an example \cite{66general}, by setting $k=3$, $4$, $5$, and $6$, four different neural network models are generated, respectively using the methods described above to predict the fidelity of the input quantum state with respect to $|\psi_0\rangle$. When estimating fidelity with a neural network model, the error of the fidelity estimation $\epsilon$ is inversely related to its confidence level $1-\delta$, where $Pr(|\tilde{F}-F|\geqslant\epsilon)\leqslant\delta$, in which $\tilde{F}$ stands for the estimated fidelity and $F$ is the actual fidelity. Figure 2 shows that the higher the number of measurement settings $k$ used, the smaller the error of the fidelity estimation $\epsilon$ is. For the same neural network model, the higher the actual fidelity $F$ is, the smaller the fidelity estimation error $\epsilon$ is.

\begin{figure}[!htbp]
  \flushleft
  \includegraphics[width=0.5\textwidth]{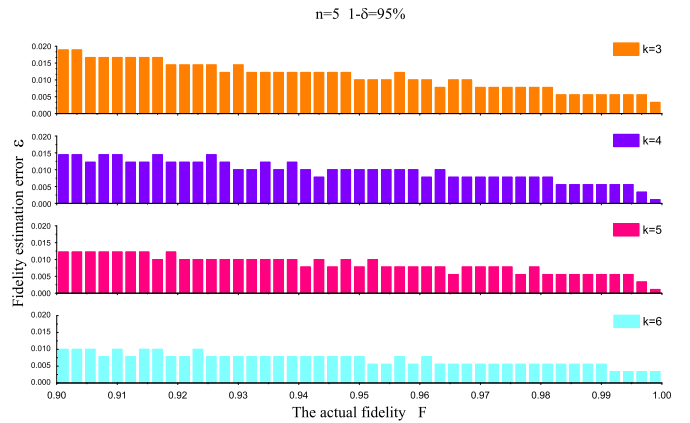}
  \caption{A plot of the fidelity estimation error $\epsilon$ of the neural network versus the actual fidelity $F$ of the quantum state when measurements are made using three, four, five, and six measurement settings. Here the target state is a five-qubit general quantum state $|\psi_0\rangle$ and the confidence level $1-\delta$ is set to be 95\%. The higher the number of measurement settings $k$ used, the smaller the error of the fidelity estimation $\epsilon$ is. The higher the actual fidelity $F$ is, the smaller the fidelity estimation error $\epsilon$ is.}
\end{figure}

\begin{figure*}[ht]
  \centering
  \subfigure[]{\includegraphics[width=0.33\textwidth]{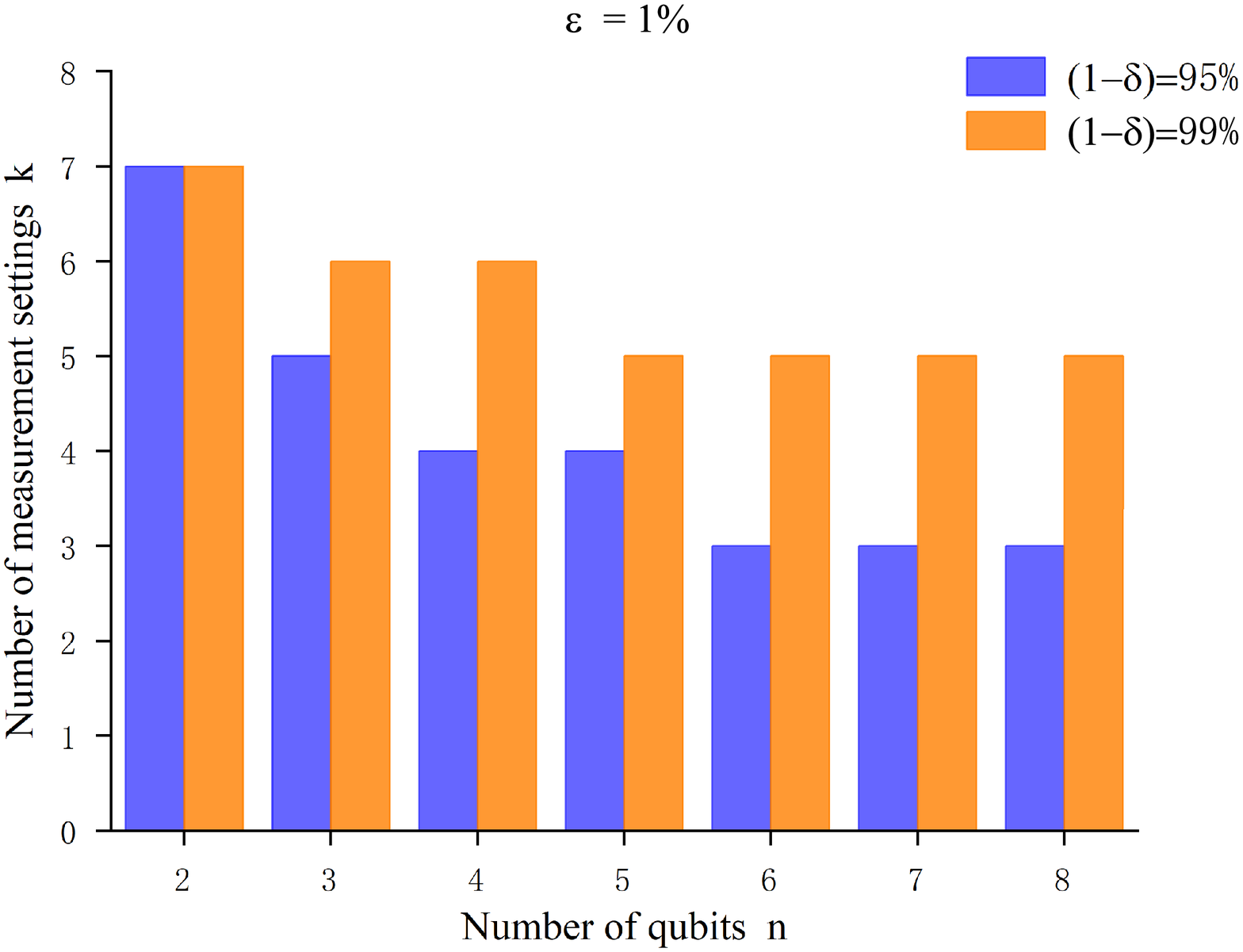}}
  \subfigure[]{\includegraphics[width=0.33\textwidth]{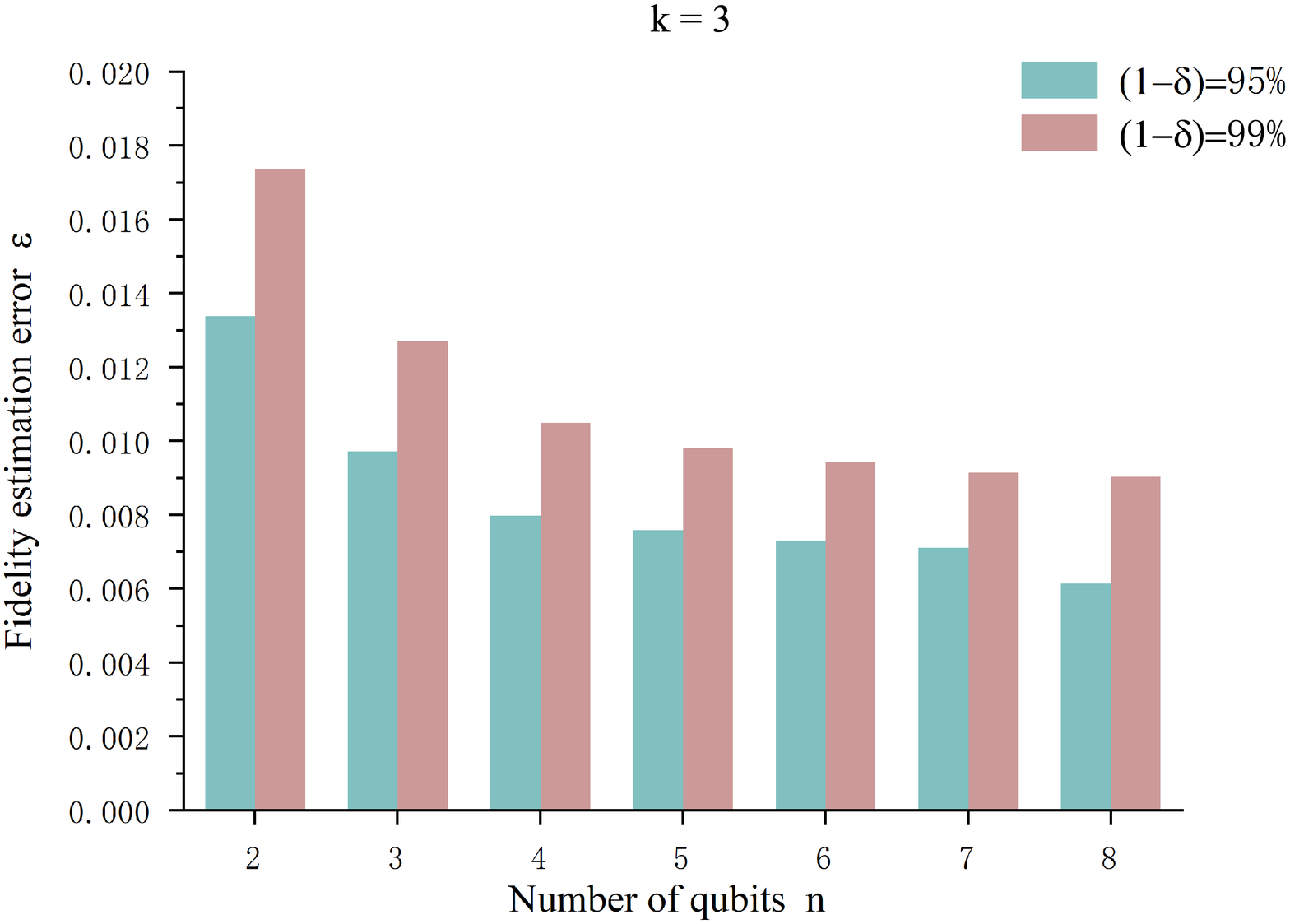}}
  \subfigure[]{\includegraphics[width=0.33\textwidth]{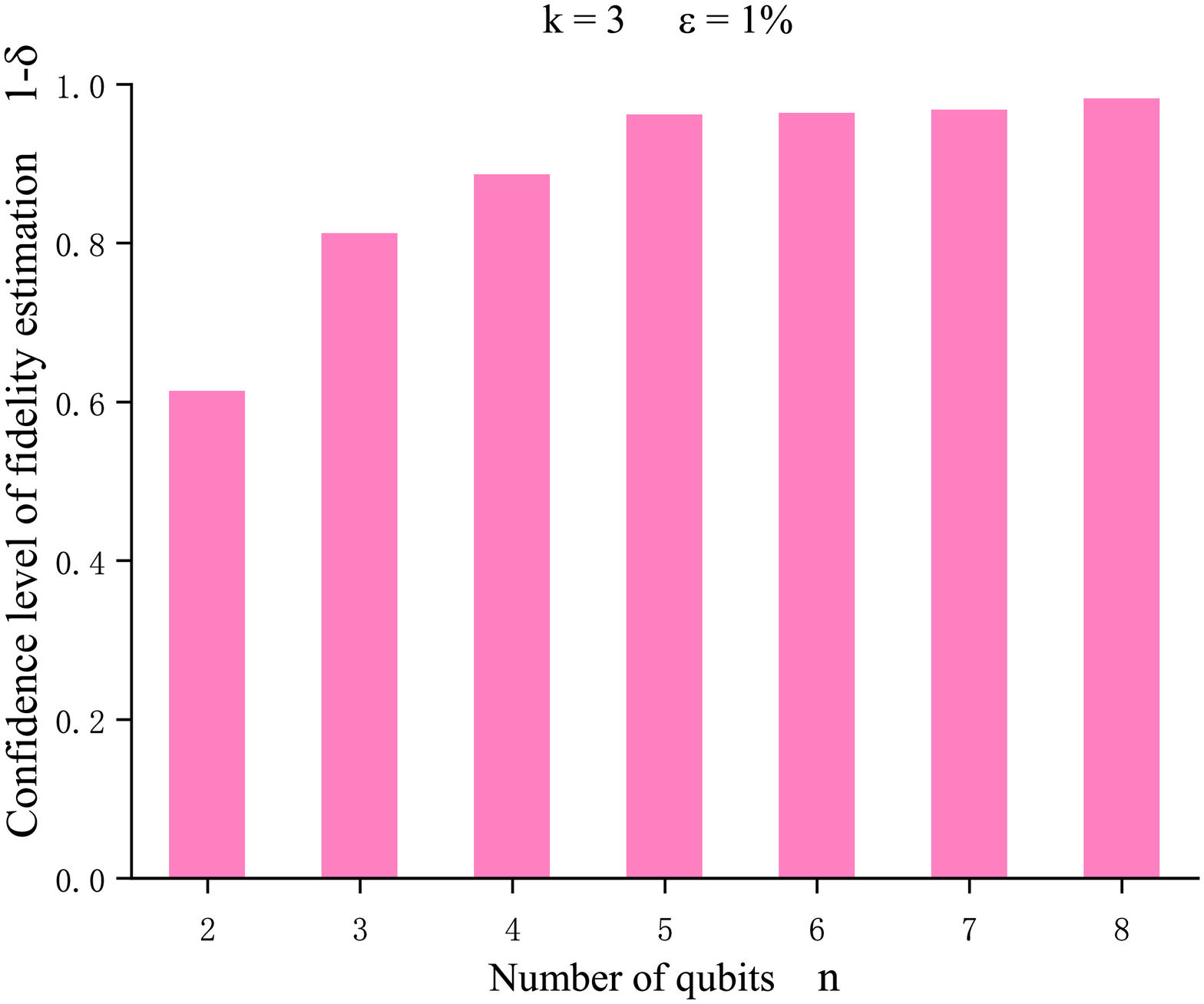}}
  \caption{The performance of this method for fidelity estimation with neural networks varies with the number of qubits in the quantum state.
  (a) The number of measurement settings $k$ decreases as the number of qubits $n$ increases when $\epsilon = 1\%$ and $1-\delta = 95\%$ (or 99\%). (b) The fidelity estimation error $\epsilon$ decreases as the number of qubits $n$ increases when $k=3$ and $1-\delta=95\%$ (or 99\%). (c) The confidence level $1-\delta$ of the fidelity estimation grows with the increase of the number of qubits $n$ when $k=3$ and $\epsilon=1\%$. We note that the fidelity of the quantum state to be estimated here is in the range of 0.95 to 1.}
\end{figure*}

Now we look at how to use a neural network model for a specific problem--to determine whether the fidelity of the input quantum states $|\psi_1\rangle$ and $|\psi_2\rangle$ \cite{70data} with respect to $|\psi_0\rangle$ exceeds 96\%. Here we choose the confidence level $1-\delta$ to be 95\%. For $|\psi_1\rangle$, we choose the top three Pauli operators in the absolute value of the expectation value for measurement, and input the measurement results into the neural network model with $k=3$, and obtain a fidelity prediction of $(97\pm 1)\%$, which indicates the fidelity of $|\psi_1\rangle$ exceeds 96\%. For $|\psi_2\rangle$, repeating the above operations, the fidelity result obtained is $(95\pm1.22)\%$, which cannot indicate whether the fidelity of $|\psi_1\rangle$ exceeds 96\% for now. Then the Pauli operators with the fourth largest absolute value of expectation value are measured, and the measurement result was input into the $k=4$ neural network model together with the measurement results of the first three Pauli operators, and the fidelity prediction result obtained is $(94.78\pm1)\%$, indicating that the fidelity of $|\psi_2\rangle$ does not exceed 96\%.

Figure 3 shows how the performance of our method for fidelity estimation with neural networks varies with the number of qubits in the quantum state. In our scheme, in addition to the number of qubits $n$, there are three other parameters, which are the number of measurement settings $k$, the fidelity estimation error $\epsilon$, and the confidence level $1-\delta$ of the fidelity estimation. Figure (3a) shows that the number of measurement settings $k$ decreases as the number of qubits $n$ increases when $\epsilon = 1\%$ and $1-\delta = 95\%$ (or 99\%). Figure (3b) shows that the fidelity estimation error $\epsilon$ decreases as the number of qubits $n$ increases when $k=3$ and $1-\delta=95\%$ (or 99\%).
Figure (3c) shows that the confidence level $1-\delta$ of the fidelity estimation grows with the increase of the number of qubits $n$ when $k=3$ and $\epsilon=1\%$. It can be seen from Fig. 3 that the performance of our method for estimating the fidelity of quantum states using neural networks is getting better as the number of qubits increases. This looks a bit counterintuitive, and we try to give an explanation for this phenomenon below.

From Eq. (3), the fidelity can be rewritten as
\begin{eqnarray}
\begin{split}
F(\rho_0,\rho)&=\sqrt{\frac{1}{2^n}\sum\limits_{j=0}^{4^n-1}\beta_j a_j}\\
&=\sqrt{\sum\limits_{j=0}^{4^n-1}P_j \gamma_j},
\end{split}
\end{eqnarray}
where $\gamma_j=\beta_j/a_j$, $P_j={a_j}^2/{2^n}$ and $\sum_{j=0}^{4^n-1}P_j=1$.
It can be seen that the fidelity $F$ is equal to the square root of the weighted average of $\gamma_j$ in which $j$ ranges from 0 to $4^n-1$.
For an ideal quantum state with unity fidelity, all $\gamma_j$ will have a value of 1; for a nonideal quantum state, these $4^n$ $\gamma_j$ will deviate from the value of 1. In the nonadversarial case, the deviation of $\gamma_j$ with respect to 1 will be general and random over all Pauli operators, and not concentrated on a few specific Pauli operators.
Therefore, a fraction of $\gamma_j$ can be selected for measurement to estimate the overall deviation of $\gamma_j$ with respect to 1 and, thus, achieve the fidelity estimation.
The larger the number of selected $\gamma_j$ (this number is defined as $l$), the smaller the error in the fidelity estimation.
When $n$ is large, $l$ does not need to increase continuously to provide a good estimate of the fidelity \cite{21Steven}. As a visual example, for the United States presidential election with 260 million voters, the sample size of the poll only needs to be around 1000 \cite{69US}.

With a fixed value of $l$, the number of required measurement settings $k$ decreases as the number of qubits $n$ increases.
This is because, for each measurement setting for an $n$-qubit quantum state, $2^n$ possible outcomes will be produced, and using them the $\gamma_j$ corresponding to $2^n-1$ nontrivial Pauli operators can be calculated.
For example, for the measurement results of one measurement setting for a three-qubit quantum state, the
$\gamma_j$ corresponding to seven nontrivial Pauli operators can be calculated, while for the measurement results of one measurement setting for a six-qubit quantum state, the $\gamma_j$ corresponding to 63 nontrivial Pauli operators can be calculated.
Thus, the number of measurement settings required to obtain $l$ $\gamma_j$ for a six-qubit state is less than that for a three-qubit state--this qualitatively explains why the performance of our method for estimating the fidelity of quantum states using neural networks is getting better as the number of qubits $n$ increases.

In the following, we compare our machine-learning-based fidelity estimation method with the other three methods QST, QSV and DFE, pointing out their respective advantages, disadvantages, and applicability.
The advantage of QST is that it can characterize a completely unknown quantum state and reconstruct its density matrix to obtain all the information, while the other three methods can only evaluate the fidelity of the input quantum state with respect to a specific target quantum state. Its drawback is that the resources consumed grow exponentially with the size of the system.
The advantage of QSV is that it can evaluate the fidelity of a quantum state with minimal resources, while its disadvantage is that it is currently only applicable to special quantum states such as stabilizer states and not to the case where the target quantum state is a multiqubit arbitrary pure quantum state.
The advantage of DFE is that it is applicable to the case where the target quantum state is an arbitrary pure quantum state, and its disadvantage is that the selection of the measurement settings requires several rounds of random switching and the total number of measurement settings is relatively large.
The advantage of our method is that it is not only applicable to the case where the target quantum state is an arbitrary pure quantum state, but also requires only a small number of measurement settings, and the number of required measurement settings does not increase with the size of the system.
For example, for a seven-qubit arbitrary quantum state, our method only needs to switch the measurement settings three times to achieve a fidelity estimation within 0.01 error with $95\%$ confidence level, while to achieve the same result, QST needs to switch the measurement device $3^7=2187$ times and DFE needs to switch the measurement device ${8/[0.01^2\times(1-95\%)]}=1.6\times 10^6$ times. Note that QSV cannot evaluate the fidelity of an arbitrary seven-qubit quantum state.
The disadvantage of our method is that it has an upper limit on the fidelity estimation accuracy, which is not suitable for situations requiring extremely high fidelity estimation accuracy.
In summary, each of the four methods has its own advantages and disadvantages, which are complementary to each other and each is suitable for different application scenarios.
Our method is most suitable for evaluating whether the fidelity of a quantum state exceeds a specific value with respect to an $n$-qubit arbitrary pure quantum state.

To summarize, we present, in this Letter, a method for predicting the fidelity of quantum states using neural network models. Compared with previous methods for quantum state fidelity estimation, our method uses fewer measurement settings and works for arbitrary quantum states. Here our method is applicable to nonadversarial scenarios. It has the potential to be used in a wide variety of local quantum information applications, such as quantum computation, quantum simulation, and quantum metrology. A future research direction is to design machine-learning-based quantum state fidelity estimation schemes in adversarial scenarios.

This work was supported by the National Key Research and Development Program (Grants No. 2017YFA0305200 and No. 2016YFA0301700), the Key Research and Development Program of Guangdong Province of China (Grants No. 2018B030329001 and No. 2018B030325001), the National Natural Science Foundation of China (Grant No. 61974168). X. Zhou acknowledges support from the National Young 1000 Talents Plan. W. Luo acknowledges support from the National Natural Science Foundation of China (Grant No. 61877029). X. Zhang acknowledges support from the National Natural Science Foundation of China (Grant No. 62005321). S. Pang acknowledges support from the National Natural Science Foundation of China (Grant No. 12075323). \\

X. Zhang and M. Luo contributed equally to this work.

%
%

\section*{Supplementary materials for ``Direct Fidelity Estimation of Quantum States using Machine Learning"}

In this supplementary material, we discuss more results. Sec.(I) describes the basic structure of artificial neural networks (ANNs) and our artificial neural network. Sec.(II) gives the method to generate quantum states with specified fidelity. It also analyzes the uniformity of pure state fidelity, mixed state fidelity The purity distribution of mixed states, the generality of our method, and Poisson noise are also presented here. The method for selecting the Pauli operators is given in Sec.(III). Sec.(IV) analyzes the accuracy of the quantum state fidelity estimation. Sec.(V) gives the verification accuracy of the neural network. Sec.(VI) presents a practical application of our neural network. Sec.(VII), we discuss the scalability of the neuron number.

\section{The structure of ANNs and our ANN}
\centerline{\textbf{The basic structure of ANNs}}

ANNs consists of an input layer, hidden layers and an output layer (See Fig.1). The input layer consists of $k\times2^n$ neurons corresponding to the probability of each outcome, in which $n$ represents the number of qubits and $k$ represents the number of Pauli combinations. We set the inputs $\bf{x_0}$ and the intermediate vector $\bf{x_1}$ in the hidden layer generated by the non-linear relation
\begin{eqnarray}
\bf{x_1}=\sigma_{RL}(W_1\bf{x_0}+\bf{\omega_1}),
\end{eqnarray}
where $\sigma_{RL}$ is the ReLU function for each neuron in the hidden layer, defined as $\sigma_{RL}(z_i)=max(z_i,0) (i=1,2,3,...)$. The matrix $\bf{W_1}$ is the initialized weight and the vector $\bf{\omega_1}$ is the bias between the input layer and the hidden layer. The optimal output vector denoted as $\bf{x_2}$ is generated using the function
\begin{eqnarray}
\bf{x_2}=\sigma_s(W_2\bf{x_1}+\bf{\omega_2}),
\end{eqnarray}
where $\sigma_s$ is the Softmax function defined by $\sigma(z_i)=\frac{e^{z_i}}{\sum_{k=1}^{122}e^{z_k}} (i=1,...,122)$. The matrix $\bf{W_2}$ is the initialized weight between the hidden layer and the output layer, while the vector $\bf{\omega_2}$ is the bias. The loss function is categorical cross-entropy and is written as $-\frac{1}{n}[y_slog a_s+(1-y_s)log(1-a_s)]$. The subscript $s$ denotes the sequence number of the training sample, and the notation $y$ represents the labels defined by the criterion, $a$ means the output labels of the ANN and $n$ is the training set number, respectively. During the machine learning process, $W_1$, $\omega_1$, $W_2$, $\omega_2$ are continuously optimized until the confidence level reaches saturation, and then the training is stopped.

\begin{figure}[!h]
  \centering
  \includegraphics[width=0.35\textwidth]{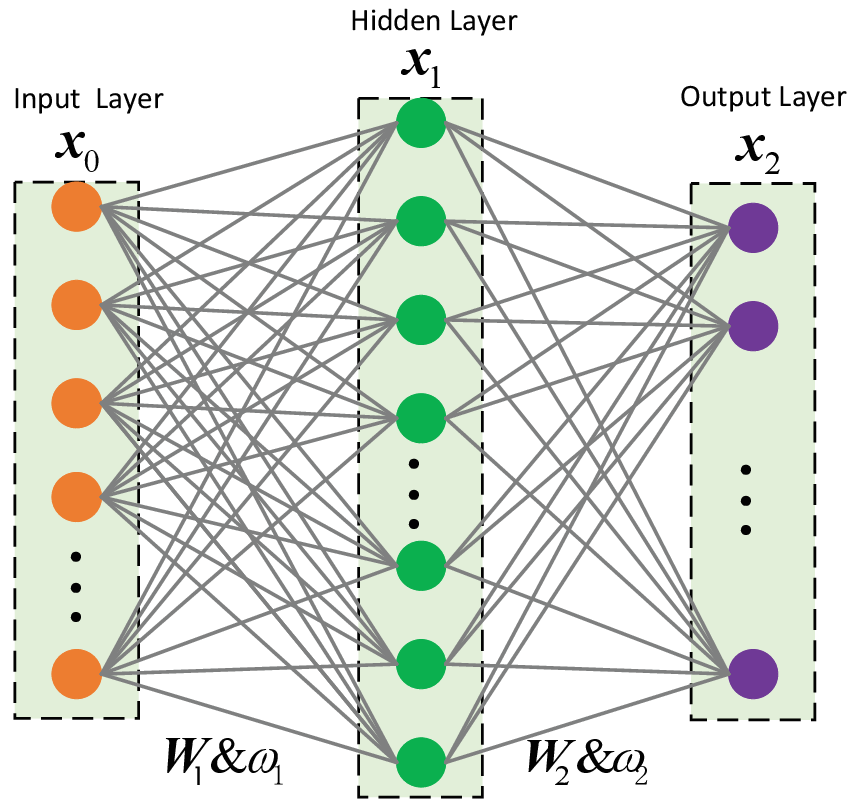}
  \caption{(Color online) Artificial neural networks with hidden layers. The objective of machine learning is to optimize $\sigma_s(\bf{W_2}\sigma_{RL}(\bf{W_1}\bf{x_0}+\bf{\omega_1})+\bf{\omega_2})$, where $\sigma_{RL}$ is ReLU function and $\sigma_s(z)_i=\frac{e^{z_i}}{\sum_{k=1}^ke^{z_k}}$ $(i=1,2,...,K)$ is softmax function. The input data $\bf{x_0}=\{a_1, a_2, \ldots, a_k\}$ are the measurements of Pauli operators. Matrix $\bf{W_1}$ and the vector $\bf{\omega_1}$ are initialized uniformly and optimized during the learning process. The number of neurons in the hidden layer can be varied for optimal performance. The output data $\bf{x_2}$ are the predicted fidelity.}\label{A1}
\end{figure}

\centerline{\textbf{The ANN for fidelity estimation}}

We use four 2080Ti GPUs. We choose the optimizer that has the best performance in our task among almost all the built-in optimizers in TensorFlow: NadamOptimizer (adaptive moment estimation). This neural network contains 122 labels, using 1,952,000 data for training and 488,000 data for validation, i.e. each label contains 16,000 training data and 4,000 validation data. By tuning the batch size of inputs, the number of neurons, and the number of training rounds, the performance of the neural network is continuously optimized. Eventually, the parameters required for training this neural network from two-qubit to seven-qubit quantum states are shown in Table S1 and Table S2. Hid-neu represents the number of the hidden neurons.

\begin{table}[!htbp]
  \begin{center}
  \begin{tabular}{|c|c|c|c|c|c|c|c|}
  \multicolumn{8}{c}{\rule[-3mm]{0mm}{5mm} Table S1. Parameters of the neural network with $k\times[\frac{(n^4-2n^3+11n^2+14n)}{24}+1]$}\\
  \multicolumn{8}{c}{\rule[-3mm]{0mm}{5mm} \qquad\qquad expected values of non-trivial Pauli operators as the neuron inputs.}\\
  \hline
  \multicolumn{4}{|c|}{four-qubit states}                                                    &\multicolumn{4}{c|}{five-qubit states}                 \\
  \hline
  state                               & epoch              & Batch size          &Hid-neu    &state        & epoch              & Batch size           & Hid-neu  \\
  \hline
  \multirow{3}{*}{$|\varphi_4\rangle$}&\multirow{3}{*}{400}&\multirow{3}{*}{8192}&2000&\multirow{3}{*}{$|\varphi_5\rangle$}&\multirow{3}{*}{500}&\multirow{3}{*}{16384}&700-300\\
                                      &                    &                     &3000&                   &                          &                           &900-300\\
                                      &                    &                     &5000&                   &                          &                           &1000-300\\
  \hline
  \multicolumn{4}{|c|}{six-qubit states}                                                   &&&&\\
  \hline
  state                               & epoch             & Batch size          &Hid-neu  &&&&\\
  \hline
  \multirow{3}{*}{$|\varphi_6\rangle$}&\multirow{3}{*}{500}&\multirow{3}{*}{16384}&500-300  &&&&\\
                                     &                    &                      &700-300  &&&&\\
                                     &                    &                      &900-300  &&&&\\
  \hline
  \end{tabular}
  \end{center}
\end{table}

\begin{table}[!htbp]
  \begin{center}
  \begin{tabular}{|c|c|c|c|c|c|c|c|}
  \multicolumn{8}{c}{\rule[-3mm]{0mm}{5mm} Table S2. Parameters of the neural network with all measurement}\\
  \multicolumn{8}{c}{\rule[-3mm]{0mm}{5mm} outcome probabilities as neuron inputs}\\
  \hline
  \multicolumn{4}{|c|}{two-qubit states}                                             &  \multicolumn{4}{c|}{three-qubit states} \\
  \hline
  state               & epoch                   & Batch size              & Hid-neu  & state                & epoch                & Batch size             & Hid-neu\\
  \hline
  Bell                & \multirow{2}{*}{200}    & \multirow{2}{*}{2048}   & 1000     & GHZ                  & \multirow{2}{*}{400} & \multirow{2}{*}{4096}  & 1000\\
  $W$                 &                         &                         & 2000     & W                    &                      &                        & 2000\\
  $|\varphi_2\rangle$ &                         &                         & 3000     & $|\varphi_3\rangle$  &                      &                        & 3000\\
  \hline
   \multicolumn{4}{|c|}{four-qubit states}                                           &  \multicolumn{4}{c|}{five-qubit states} \\
  \hline
  state               & epoch                   & Batch size              & Hid-neu  & state                & epoch                & Batch size             & Hid-neu\\
  \hline
  Cluster             & \multirow{5}{*}{400}    & \multirow{5}{*}{8192}   &          & Cluster              & \multirow{6}{*}{200} & \multirow{6}{*}{16384} & \\
  W                   &                         &                         & 2000     & C-ring               &                      &                        & 500-500\\
  GHZ                 &                         &                         & 3000     & Dicke                &                      &                        & 1000-1000\\
  Dicke               &                         &                         & 5000     & GHZ                  &                      &                        & 1500-1500\\
  $|\varphi_4\rangle$ &                         &                         &          & W                    &                      &                        & \\
  \cline{1-4}
  \multicolumn{4}{|c|}{six-qubit states}                                                      & $|\varphi_5\rangle$  &                      &                        & \\
  \hline
  state               & epoch                   & Batch size              & Hid-neu  & \multicolumn{4}{c|}{seven-qubit states}    \\
  \hline
  $C_{23}$            & \multirow{5}{*}{500}    & \multirow{5}{*}{16384}  &          & state                & epoch                & Batch size             & Hid-neu\\
  \cline{5-8}
  Dicke               &                         &                         & 500-500  & \multirow{4}{*}{$|\varphi_7\rangle$}       & \multirow{4}{*}{500} & \multirow{4}{*}{16384}&\\
  GHZ                 &                         &                         & 1000-1000&              &                      &                        & 800-400\\
  W                   &                         &                         & 1500-1500&              &                      &                        & 900-400\\
  $|\varphi_8\rangle$ &                         &                         &          &              &                      &                        & 1000-400\\
  \hline
   \multicolumn{4}{|c|}{eight-qubit states}                                          &     &&& \\
  \hline
  state               & epoch                   & Batch size              & Hid-neu  &     &&&    \\
  \hline
  \multirow{3}{*}{$|\varphi_8\rangle$} & \multirow{3}{*}{500}    & \multirow{3}{*}{16384}  &700-400   &     &&& \\
  \cline{5-8}
                      &                         &                         &1000-400  &     &&&\\
                      &                         &                         & 2000-500 &     &&& \\
  \hline
  \end{tabular}
  \end{center}
\end{table}

Next, we show the specific forms of the special number states that appear in Table S2 (See Fig.2). Moreover, $|\varphi_2\rangle$, $|\varphi_3\rangle$, $|\varphi_4\rangle$, $|\varphi_5\rangle$, $|\varphi_6\rangle$, $|\varphi_7\rangle$ and $|\varphi_8\rangle$ are general quantum states \cite{5states}.
\noindent The two-qubit Bell state and the W state are
\begin{eqnarray}
\begin{array}{l}
\displaystyle |\phi^+\rangle_{Bell}=\frac{1}{\sqrt{2}}(|00\rangle+|11\rangle), \\
\displaystyle |W\rangle=\frac{1}{\sqrt{3}}(|00\rangle+|01\rangle+|10\rangle).
\end{array}
\end{eqnarray}

\noindent The three-qubit GHZ state and the W state are
\begin{eqnarray}
\begin{array}{l}
\displaystyle |GHZ\rangle=\frac{1}{\sqrt{2}}(|000\rangle+|111\rangle),\\
\displaystyle |W\rangle=\frac{1}{\sqrt{3}}(|001\rangle+|010\rangle+|100\rangle).
\end{array}
\end{eqnarray}

\noindent The four-qubit Cluster state, the Dicke state, the GHZ state and the W state are
\begin{eqnarray}
\begin{array}{l}
\displaystyle |Cluster_4\rangle=\frac{1}{2}(|0000\rangle+|0011\rangle+|1100\rangle-|1111\rangle),\\
\displaystyle |Dicke_4^2\rangle=\frac{1}{\sqrt{6}}(|0011\rangle+|0110\rangle+|0101\rangle+|1010\rangle\\
\displaystyle \qquad\qquad\qquad+|1001\rangle+|1100\rangle),\\
\displaystyle |GHZ\rangle=\frac{1}{\sqrt{2}}(|0000\rangle+|1111\rangle),\\
\displaystyle |W\rangle=\frac{1}{2}(|0001\rangle+|0010\rangle+|0100\rangle+|1000\rangle).
\end{array}
\end{eqnarray}

\noindent The five-qubit Cluster state, the C-ring state, the Dicke state, the GHZ state and the W state are
\begin{eqnarray}
\begin{array}{l}
\displaystyle \qquad|Cluster_5\rangle=\frac{1}{2}(|+0+0+\rangle+|+0-1-\rangle\\
\displaystyle \qquad\qquad\qquad+|-1-0+\rangle+|-1+1-\rangle),\\
\displaystyle \qquad|C\text{-}ring_5\rangle=\frac{1}{2\sqrt{2}}(|+0+00\rangle+|-0+01\rangle\\
\displaystyle \qquad\qquad\qquad +|+0-10\rangle-|-0-11\rangle+|-1-00\rangle\\
\displaystyle \qquad\qquad\qquad +|+1-01\rangle+|-1+10\rangle-|+1+11\rangle),\\
\displaystyle \qquad|Dicke_5^2\rangle=\frac{1}{\sqrt{10}}(|11000\rangle)+|10100\rangle+|01100\rangle\\
\displaystyle \qquad\qquad\qquad +|01010\rangle+|00110\rangle+|00101\rangle+|00011\rangle\\
\displaystyle \qquad\qquad\qquad +|10010\rangle+|10001\rangle+|01001\rangle),\\
\displaystyle \qquad|GHZ\rangle=\frac{1}{\sqrt{2}}(|00000\rangle+|11111\rangle),\\
\displaystyle \qquad|W\rangle=\frac{1}{\sqrt{5}}(|10000\rangle+|01000\rangle+|00100\rangle\\
\displaystyle \qquad\qquad\qquad +|00010\rangle+|00001\rangle).
\end{array}
\end{eqnarray}

\noindent The six-qubit Cluster state, the Dicke state, the GHZ state and the W state are
\begin{eqnarray}
\begin{array}{l}
\displaystyle \qquad|C23\rangle=\frac{1}{2}(|+0++0+\rangle+|+0+-1-\rangle\\
\displaystyle \qquad\qquad\qquad+|-1-+0+\rangle-|-1--1-\rangle),\\
\displaystyle \qquad|Dicke_6^2\rangle=\frac{1}{\sqrt{15}}(|000011\rangle+|000101\rangle+|001010\rangle\\
\displaystyle \qquad\qquad\qquad+|001100\rangle+|110000\rangle+|010100\rangle+|101000\rangle\\
\displaystyle \qquad\qquad\qquad +|001001\rangle+|010001\rangle+|100001\rangle +|100010\rangle\\
\displaystyle \qquad\qquad\qquad+|100100\rangle+|000110\rangle+|011000\rangle+|010010\rangle),\\
\displaystyle \qquad|GHZ\rangle=\frac{1}{\sqrt{2}}(|000000\rangle+|111111\rangle),\\
\displaystyle \qquad|W\rangle=\frac{1}{\sqrt{6}}(|000001\rangle+|000010\rangle+|000100\rangle\\
\displaystyle \qquad\qquad+|010000\rangle+|100000\rangle.
\end{array}
\end{eqnarray}

\begin{figure*}[!htb]
  \flushleft
  \includegraphics[width=0.4\textwidth]{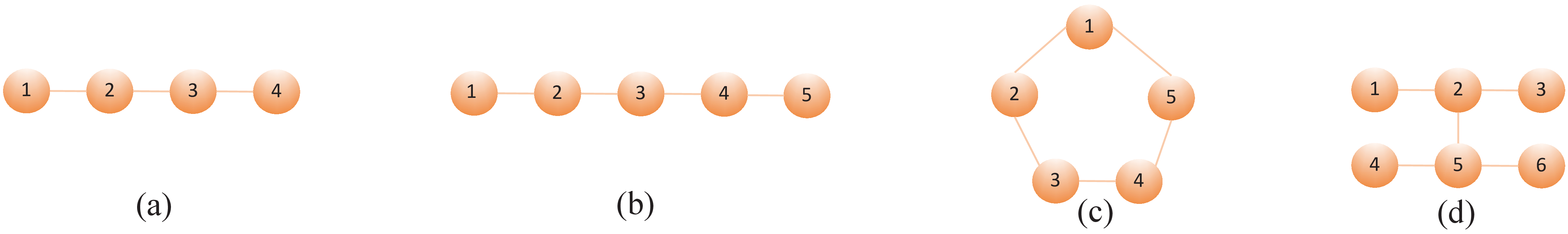}
  \caption{The structure of Cluster states. (a) The four-qubit line Cluster state $|Cluster_4\rangle$. (b) The five-qubit line Cluster state $|Cluster_5\rangle$. (c) The five-qubit ring Cluster state $|C\text{-}ring\rangle$. (d) The six-qubit grid Cluster state $|C23\rangle$.}\label{A1}
\end{figure*}

\section{Generation and analysis of quantum states}
\centerline{\textbf{A. Generation of quantum pure states with specified fidelity}}

Preparing quantum states has an important role in realizing quantum information and quantum computing, but often the imperfection of devices and the influence of noise result in obtaining quantum states that are all mixed states. Here, we use neural network techniques to evaluate whether the fidelity between this quantum state and the ideal state satisfies the requirements. Our neural network inputs are derived from mixed states with specified fidelity. We first introduce the method for generating a pure state.

Step 1. Generating an arbitrary pure state\\
In Mathematics, we create a pure state of arbitrary dimension with the help of the function \emph{RandomKet(D)} \cite{40Miszczak}. Specifically, The \emph{RandomKet(D)} function calls \emph{RandomSimplex(D)} and \emph{Randomreal(D)} to generate a D-dimensional arbitrary pure state.
The Mathematic code of generating an arbitrary pure state is shown below.

\begin{lstlisting}[frame=shadowbox]
RandomSimplex[d_]:=Block[{r,r1,r2},
r=Sort[Table[RandomReal[{0,1}], {i,1,d-1}]];
  r1=Append[r,1];
      r2=Prepend[r,0]; r1-r2
];

RandomKet[n_]:=Block[{p, ph},
p=Sqrt[RandomSimplex[n]];
ph=Exp[I RandomReal[{0,2\[Pi]},n-1]];
ph=Prepend[ph,1];
p*ph
];
\end{lstlisting}

Step 2. Generation of a pure state with specified fidelity corresponding to the state $|0\rangle^{\otimes n}$\\
An arbitrary pure state can be expanded as
\begin{eqnarray}
|\varphi\rangle=\sum_{i=0}^{2^n-1}\alpha_i|i\rangle,
\end{eqnarray}
where $|i\rangle$ $(i=0, 1, ..., 2^n-1)$ is basis vector of calculations. For convenience, we rewrite Eq.(9) as follows.
\begin{eqnarray}
|\varphi\rangle=f|0\rangle^{\otimes n}+\sum_{i=1}^{2^n-1}\alpha_i|i\rangle=f|0\rangle+\sqrt{1-f^2}|\phi\rangle^{2^n-1},
\end{eqnarray}
where $\alpha_0=f$. The state $|\phi\rangle^{2^n-1}$ is a $(2^n-1)$ dimensional arbitrary pure state. Therefore, the fidelity $f$ between $|0\rangle^{\otimes n}$ and $|\varphi\rangle$ is given as
\begin{eqnarray}
f=F(|0\rangle^{\otimes n},|\varphi\rangle)=|\langle0|^{\otimes n}|\varphi\rangle|
\end{eqnarray}
Hence, we can generate an $n$-qubit pure state dataset with the target state $|0\rangle^{\otimes n}$ using Matlab.

Step 3. Generation of a pure state with a specified fidelity corresponding to an arbitrary pure state\\
For convenience, we set $|\textbf{0}\rangle=|0\rangle^{\otimes n}$. Then we rewrite the Eq.(11) as follows.
\begin{eqnarray}
f=F(|\bf{0}\rangle,\sigma)=\sqrt{\langle\bf{0}|\sigma|\bf{0}\rangle},
\end{eqnarray}
where $\sigma$ can be viewed as a state in the dataset $S$ with a target state $|\bf{0}\rangle$. If we choose a new target pure state $\rho$, there is a unitary matrix transformation $U$ from $|\bf{0}\rangle\langle \bf{0}|$ to $\rho$. This unitary $U$ can be calculated by the code \emph{Findunitary.m}. As soon as such unitary is found \cite{4Cramer,52Tacchino}, the relative state of $\sigma$ is directly obtained as
\begin{eqnarray}
\sigma'=U\sigma U^\dagger,
\end{eqnarray}
where the state $\sigma'$  belongs to the database $S'$ of the target state $\rho$.  The fidelity $f'$ between the target state $\rho$ and the state $\sigma'$ can be calculated, that is,
\begin{eqnarray}
f'=F(\rho,\sigma')=\sqrt{tr(\rho,\sigma')}=\sqrt{\langle\bf{0}|U^\dagger\sigma'U|\bf{0}\rangle}=\sqrt{\langle\bf{0}|\sigma|\bf{0}\rangle}=f.
\end{eqnarray}

Step 4. Projective measurements\\
Each measurement setting $k$ is characterized by $W_k$, and each specific result $p(k_j)$ is associated with a projection operator
\begin{eqnarray}
P_{k_j}=|v_{k_j}\rangle\langle v_{k_j}|, \quad j=1,2,...,2^n,
\end{eqnarray}
where $|v_{k_j}\rangle$ is the $j_{th}$ eigenvector of the $W_k$, $p(k_j)$ is equal to $tr(\rho P_{k_j})$.
\\

\centerline{\textbf{B. Generation of quantum mixed states with specified fidelity}}
Here we give the method for generating mixed states with specified fidelity. In a Ginibre matrix $G$, each element is the standard complex normal distribution $CN(0,1)$. The random density matrix of mixed states can be written as
\begin{eqnarray}
\rho=\frac{GG^\dagger}{tr(GG^\dagger)}.
\end{eqnarray}

Inspired by the Ginibre matrix $G$, we propose a method to prepare specified fidelity states with the target state $|\bf{0}\rangle$, i.e., the density matrix of desired N-qubit mixed state is
\begin{eqnarray}
\rho_n=\frac{\mathbb{G}_n\mathbb{G}_n^\dagger}{tr(\mathbb{G}_n\mathbb{G}_n^\dagger)}.
\end{eqnarray}
where the matrix $\mathbb{G}_n$ can be expressed as
\begin{widetext}
\begin{small}
\begin{eqnarray}
\mathbb{G}_n=\left(
  \sqrt{m_1}\left(\begin{array}{c}
  x_1e^{-2\pi i*rand_{11}}   \\
  \sqrt{1-x_1^2}e^{-2\pi i*rand_{12}}|\varphi_1\rangle \\
  \end{array} \right),...,\sqrt{m_n}\left(\begin{array}{c}
  x_ne^{-2\pi i*rand_{n1}}   \\
  \sqrt{1-x_n^2}e^{-2\pi i*rand_{n2}}|\varphi_n\rangle \\
  \end{array}\right)\right)
\end{eqnarray}
\end{small}
\end{widetext}
The notations \emph{$rand_{b1}$} and \emph{$rand_{b2}$} ($b=1,2,...,n$) represent random numbers. The set $\{|\varphi_1\rangle, ..., |\varphi_n\rangle\}$ is a collection of $2^n-1$ dimensional pure states produced by the function \emph{RandomKet}. $\{m_1, .., m_n\}$ is a set of real numbers of $2^N$ dimension normalized standard normal distribution. $\{x_1, ..., x_n\}$ is a set of undefined real numbers that are closely related to the expected fidelity.

The steps for preparing the mixed state are similar to those for preparing the pure state. As can be seen from Eq.(18), the density matrix of a mixed state with specified fidelity needs to determine the values of $\{m_1,...,m_n\}$ and $\{x_1,...,x_n\}$, where the former is generated randomly using the Matlab code and the latter is determined according to the corresponding constraints.

\emph{An example.}--For a two-qubit state, suppose the desired fidelity is $f_0$ for the target state $|00\rangle$. The values $\{x_1, x_2, x_3, x_4\}$ in the matrix $\mathbb{G}_n$ can be determined from Eqs. (18-21) using Matlab. The value range of $x_1$ can first be determined in Eq.(18) by the given fidelity $f_0$. When we fix the $x_1$ randomly and uniformly, the value range of $x_2$ can also be determined in Eq.(19). Then we fix the $x_2$ randomly and uniformly, and the value range of $x_3$ can also be defined in Eq.(20). Once the value of $x_3$ is fixed randomly and uniformly, the value of $x_4$ is also fixed in Eq.(21). Therefore, we obtain the matrix $\mathbb{G}_n$.

\begin{widetext}
\begin{small}
\begin{eqnarray}
x_1\in[min\{max(\frac{f_0^2-\sum\limits_{i=2}^4 m_i}{m_1},0),1\},min\{max(\frac{f_0^2}{m_1},0),1\}]\\
x_2\in[min\{max(\frac{f_0^2-m_1x_1^2\sum_{i=3}^4 m_i}{m_2},0),1\}, min\{max(\frac{f_0^2-m_1x_1^2}{m_1},0),1\}], \\
x_3\in[min\{max(\frac{f_0^2-\sum\limits_{i=1}^2m_ix_i^2-m_4}{m_2},0),1\}, min\{max(\frac{f_0^2-\sum\limits_{i=1}^2m_ix_i^2}{m_2},0),1\}],\\
x_4=\frac{f_0^2-\sum\limits_{i=1}^3m_ix_i^2}{m_4}.
\end{eqnarray}
\end{small}
\end{widetext}

\centerline{\textbf{C. Uniformity analysis of pure state fidelity}}

We use a computer to generate 500 single-qubit pure states. The fidelity between each of 500 pure states and the state $|0\rangle$ is $\sqrt{0.5}$. We also give a geometric representation of the Bloch ball in Fig.3(a). It is obvious that the distribution of 500 states is uniform.

\begin{figure*}[htbp]
  \centering
  \subfigure[]{\includegraphics[width=1.6in,height=1.6in]{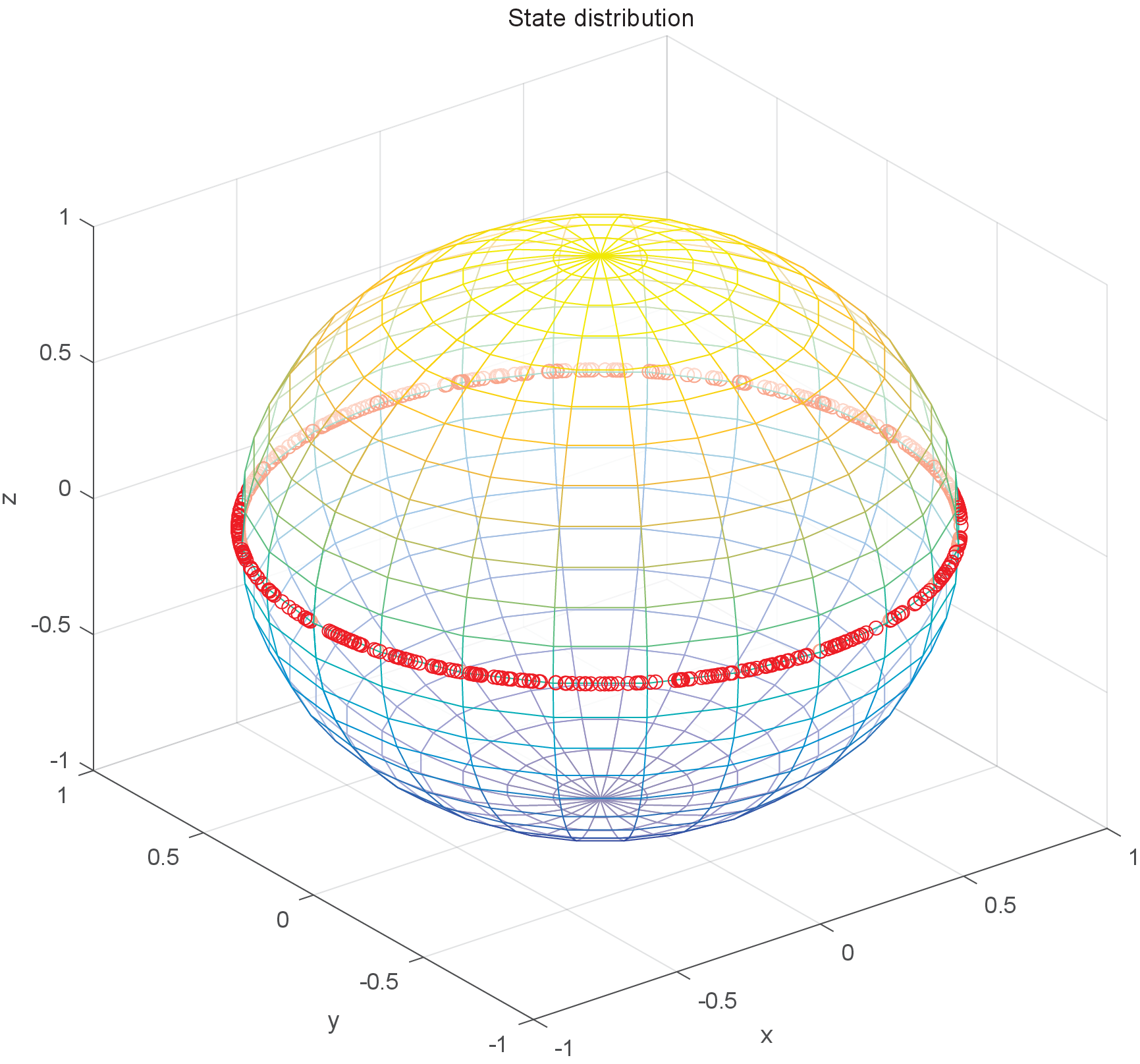}}
  \subfigure[]{\includegraphics[width=2in]{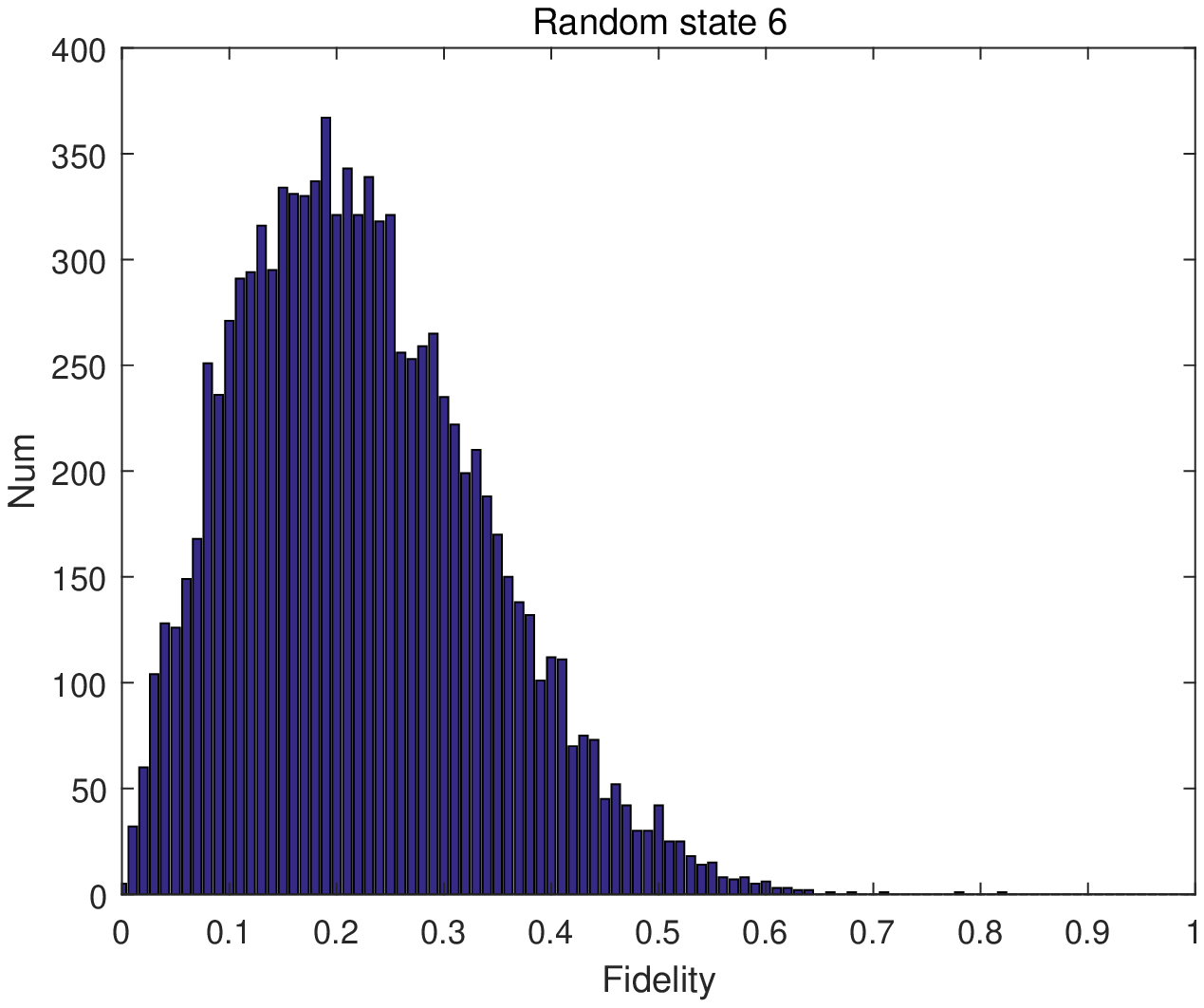}
                         \includegraphics[width=2in]{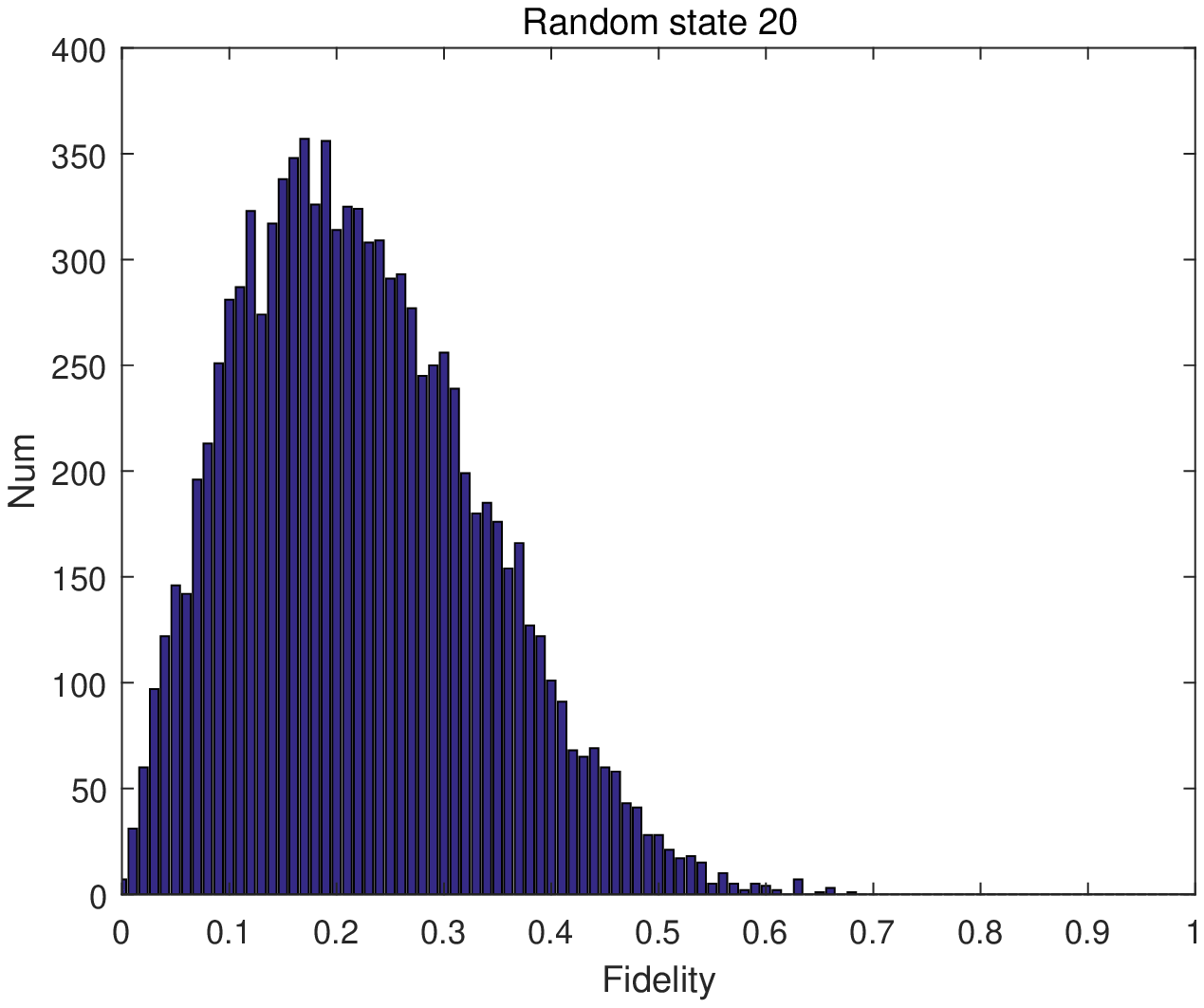}}
  \subfigure[]{\includegraphics[width=2in]{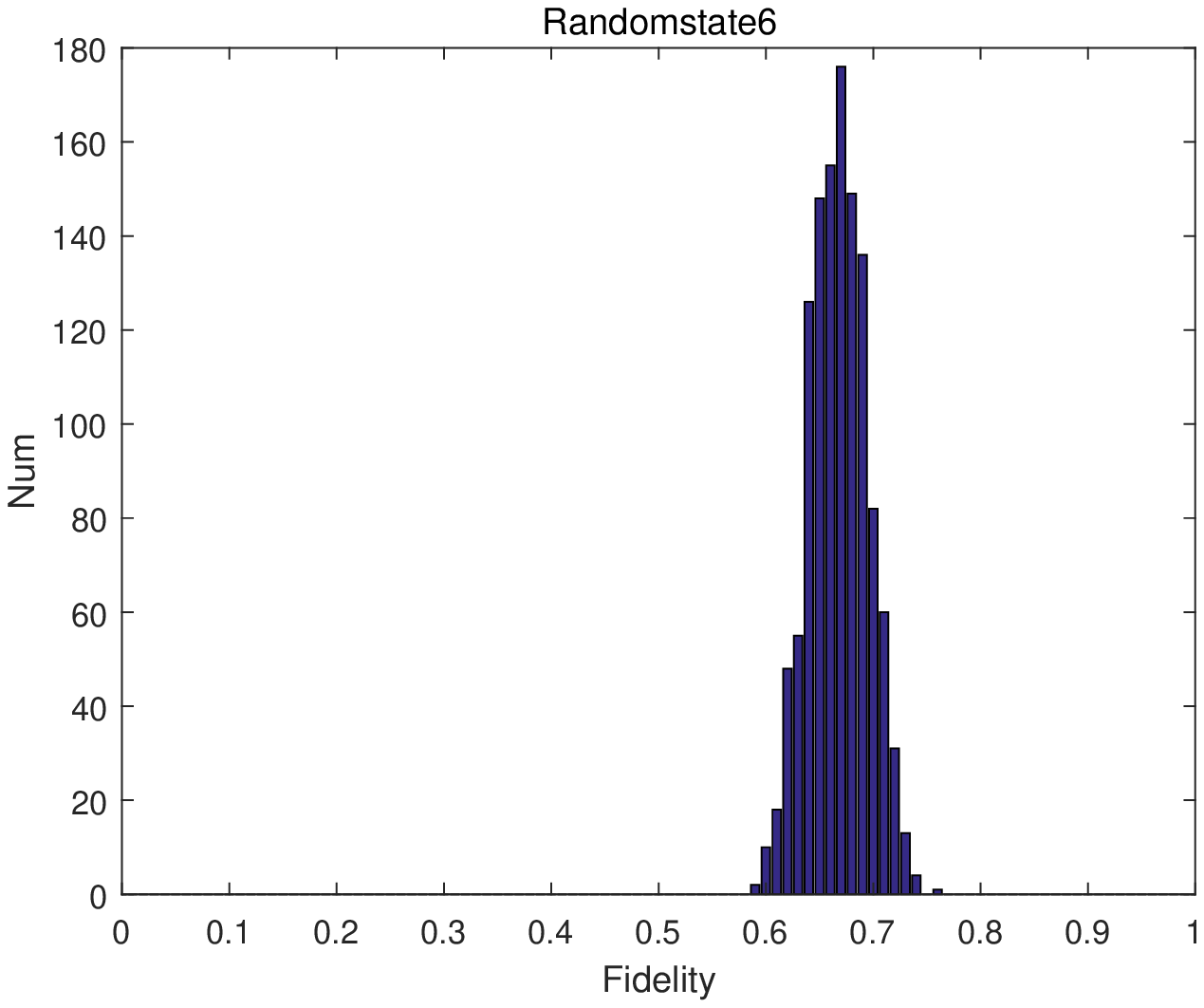}
                        \includegraphics[width=2in]{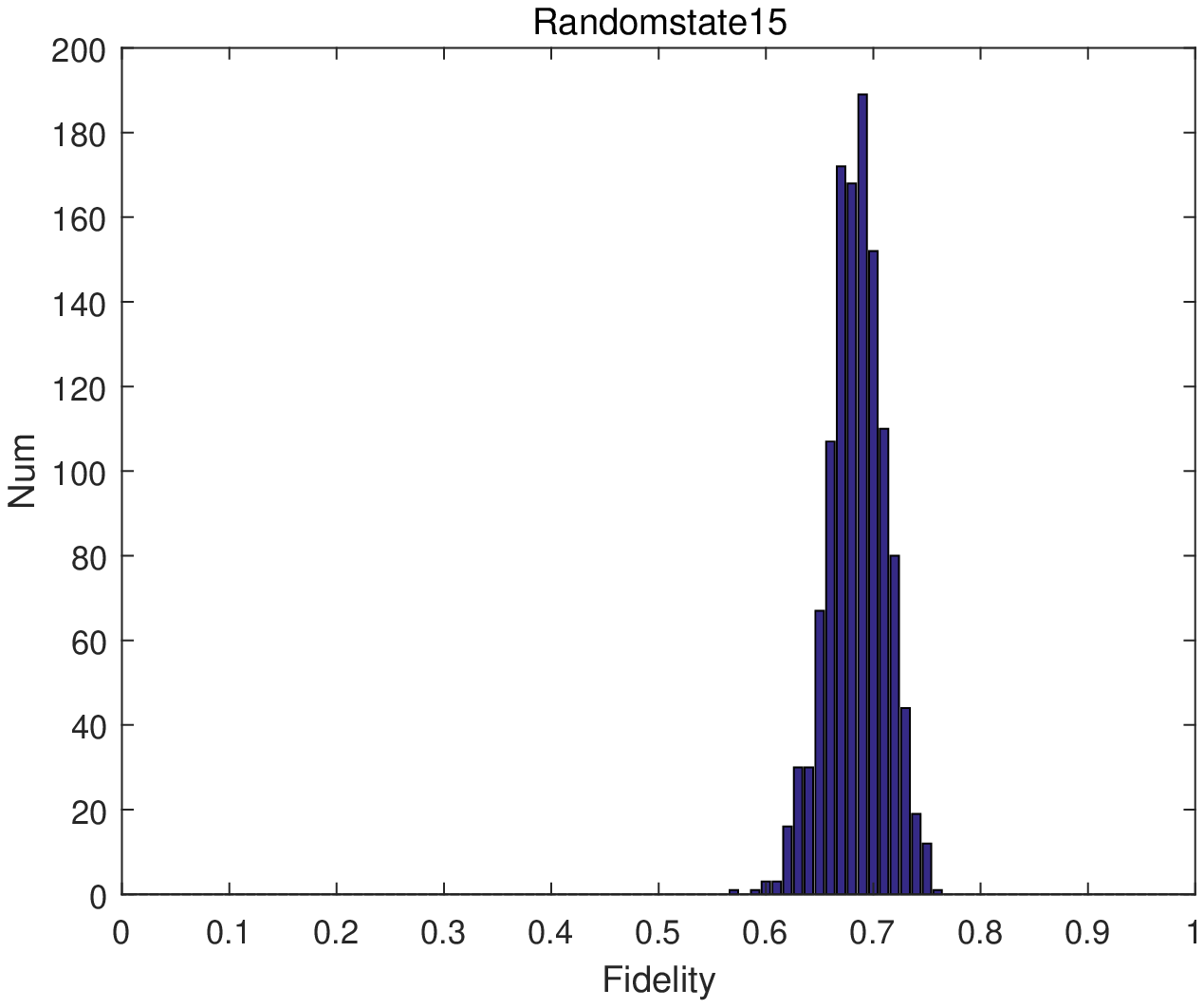}
                        \includegraphics[width=2in]{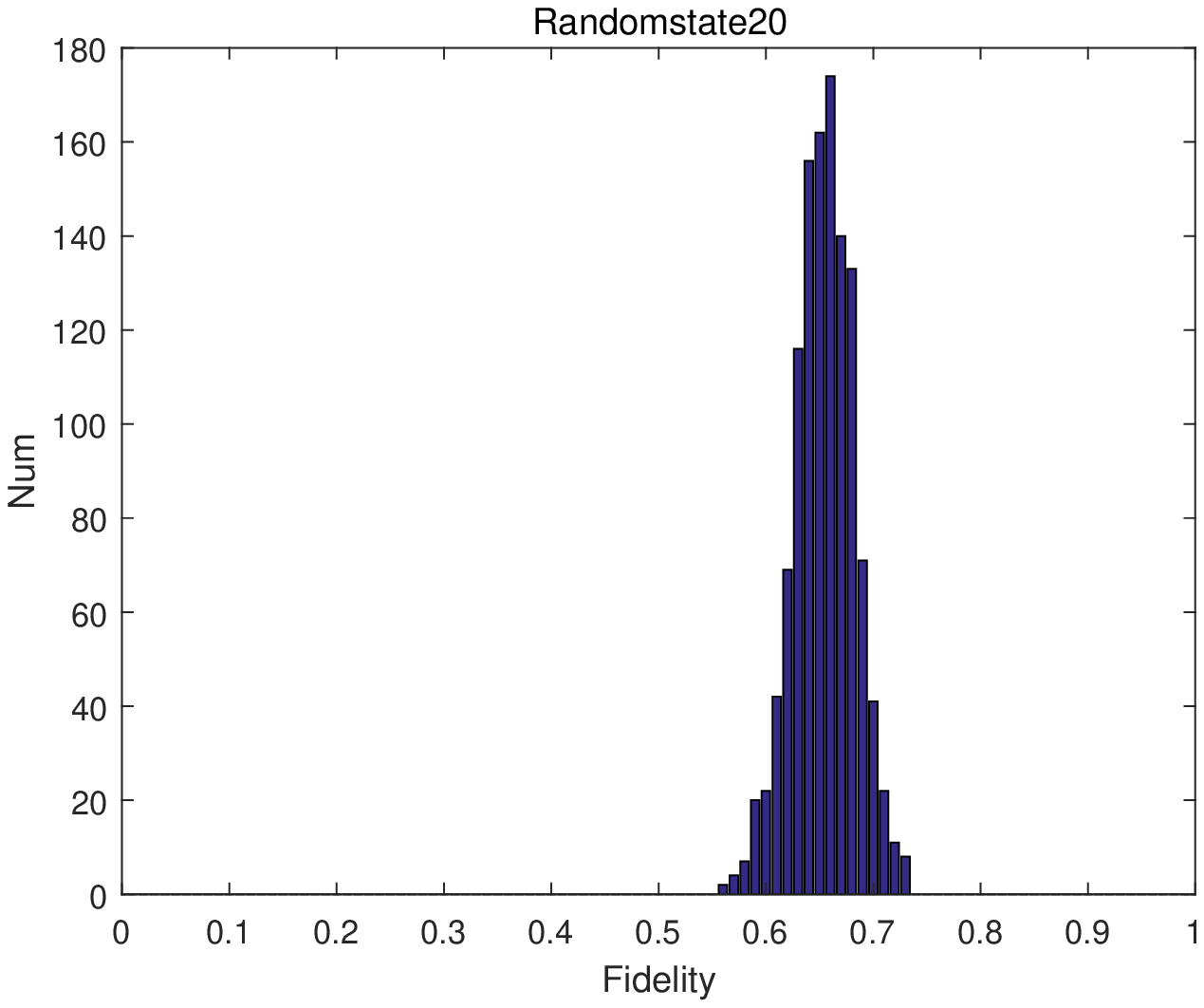}}
  \caption{(Color Online) (a) Distribution of 500 single-qubit states in the Bloch ball; (b)The distribution of 10,000 four-qubit pure states containing state 6 and
           state 20; (c) The distribution of 1,220 four-qubit mixed states including state 6, state 15 and state 20.}
\end{figure*}

In addition, we verify the uniformity of four-qubit pure states fidelities. We generate 10,000 four-qubit pure states, in which the fidelity between each of these states and the state $|0000\rangle$ is 0.25. We select 20 states out of the 10,000 states, and the fidelity between each selected state and 9,999 other states can be calculated. We get 20 similar distributions, including the distribution of random state 6 and the distribution of the random state 20 in Fig.3(b). We conclude that the distribution of 10,000 states is uniform.

\vspace{4mm}
\centerline{\textbf{D. Uniformity analysis of mixed state fidelity}}
Here, we verify the uniformity of four-qubit mixed states datasets in Fig.3(c). A total of 1,220 states are generated, in which the fidelity between 1,220 states and the state $|0000\rangle$ is 0.25. We also select 20 states out of 1,220 states. The fidelity, between each selected state and the other 1,219 states, can be calculated. We get 20 similar distribution patterns, including the distribution of random state 6, random state 15 and the distribution of random state 20 in Fig.3(c). We found that the distribution of the 1,220 states is uniform.

\vspace{4mm}
\centerline{\textbf{E. Distribution of different purities of mixed states}}
The purity of a quantum state $\rho$ is defined as $tr(\rho^2)$, where $tr(\rho^2)=1$ means that this quantum state is pure and vice versa is a mixed state. Figure 4 shows the distribution of purity for 1,220 four-qubit mixed states, in which $m_1$ can control the purity of quantum states. In Fig.4(a-b) the fidelity $f_0$ between the prepared state and $|0000\rangle$ is 0.25, and the fidelity $f_1$ is 0.8 in Fig.4(c-d). The controller $m_1$ is equal to 1, 0.9, 0.6, 0.2 or 0.01. Moveover, $m_1$ can also be a uniform distribution $U(0,1)$. The analysis here demonstrates how to control the purity of the prepared states.

\begin{figure*}[htbp]
  \centering
  \subfigure[$f_0$=0.25]{\includegraphics[width=2in]{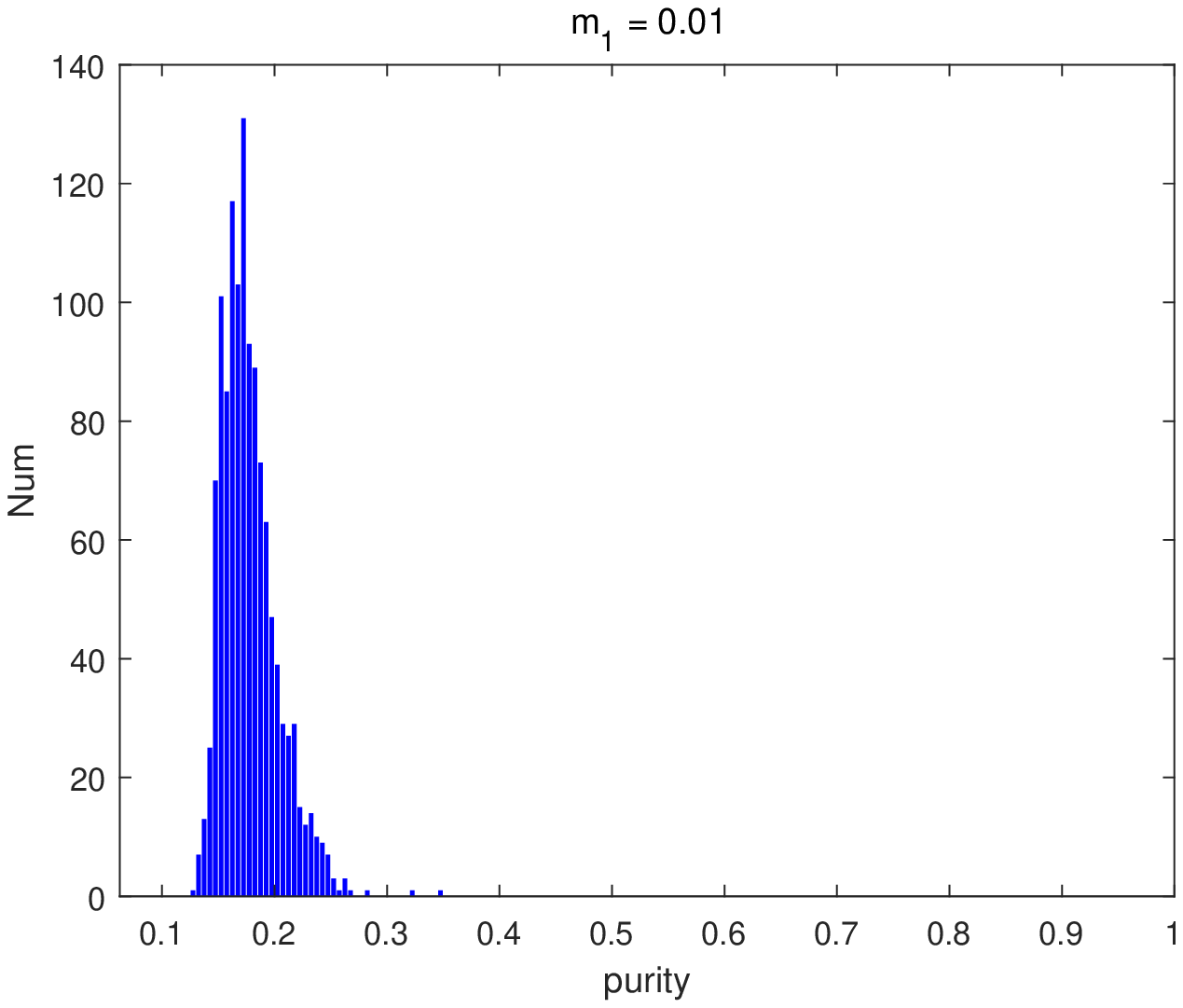}
               \includegraphics[width=2in]{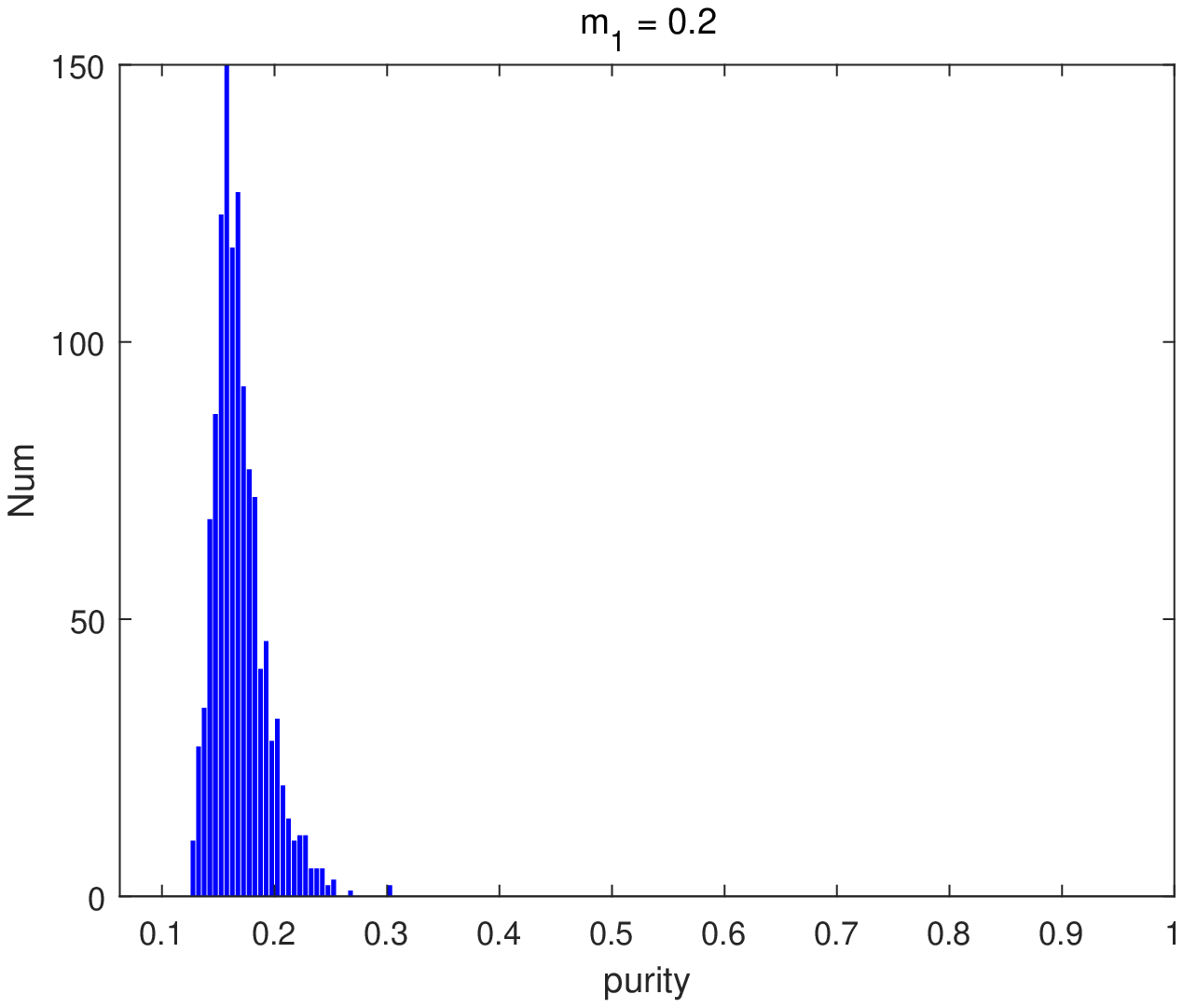}
               \includegraphics[width=2in]{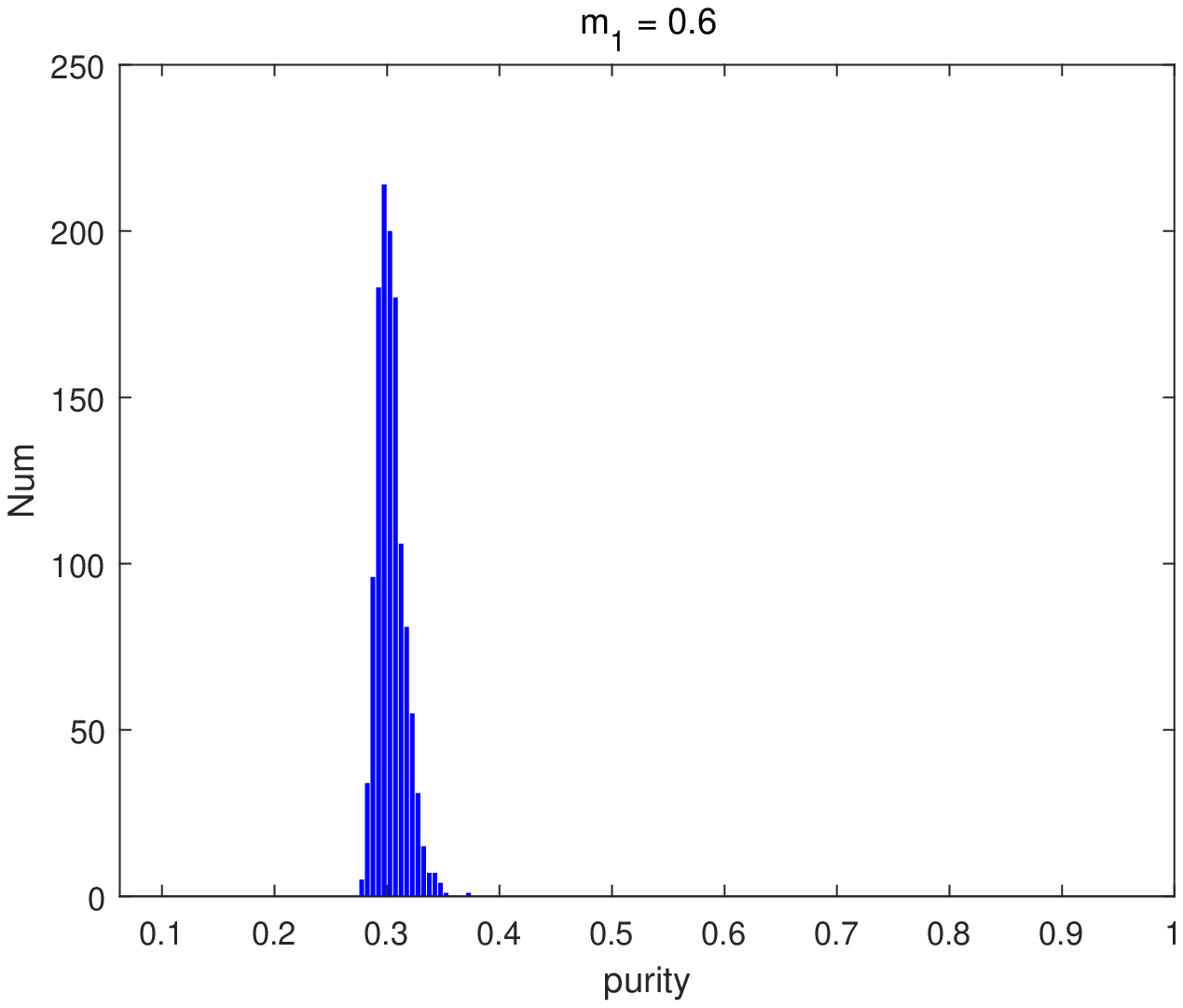}}
  \subfigure[$f_0$=0.25]{\includegraphics[width=2in]{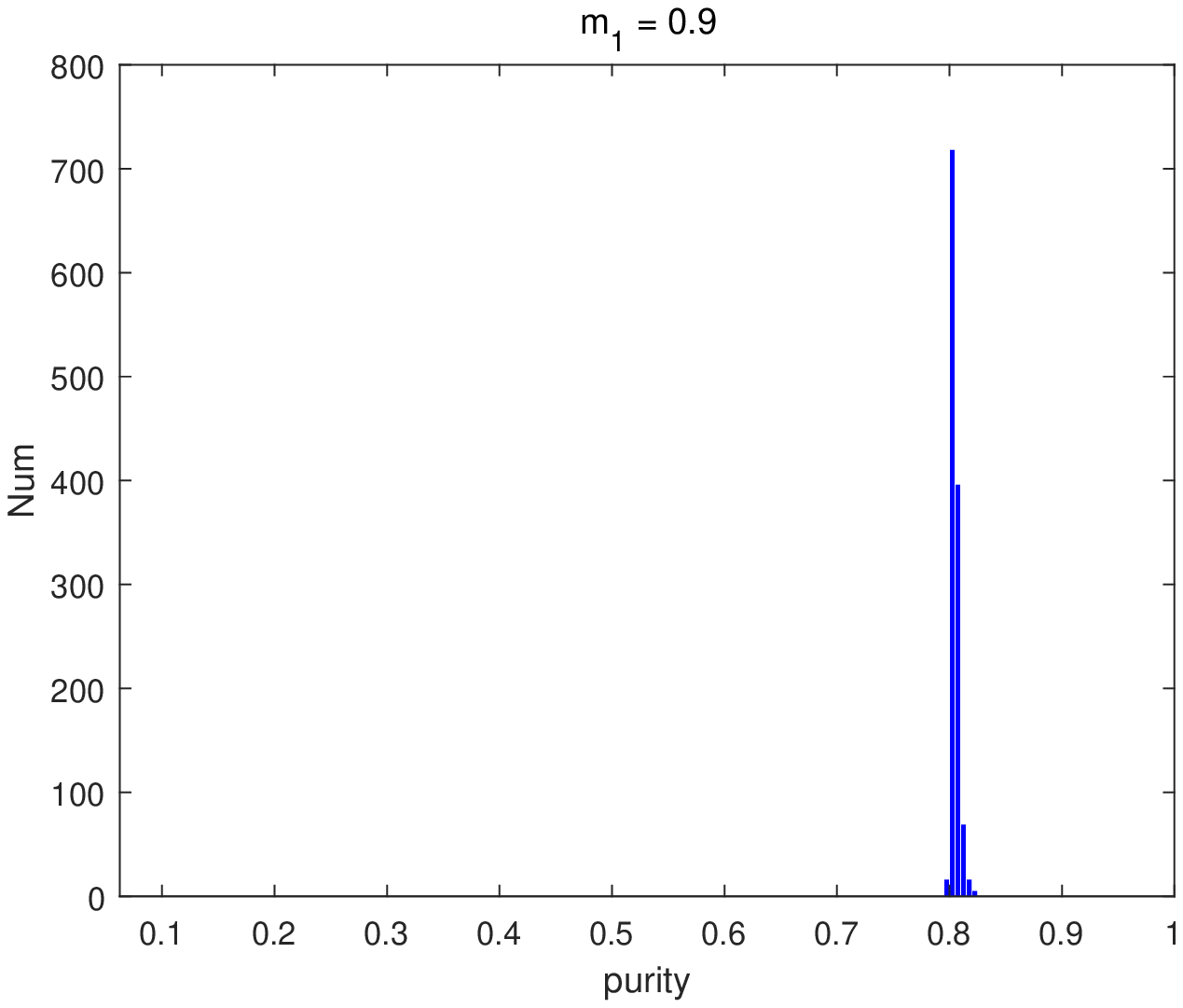}
                        \includegraphics[width=2in]{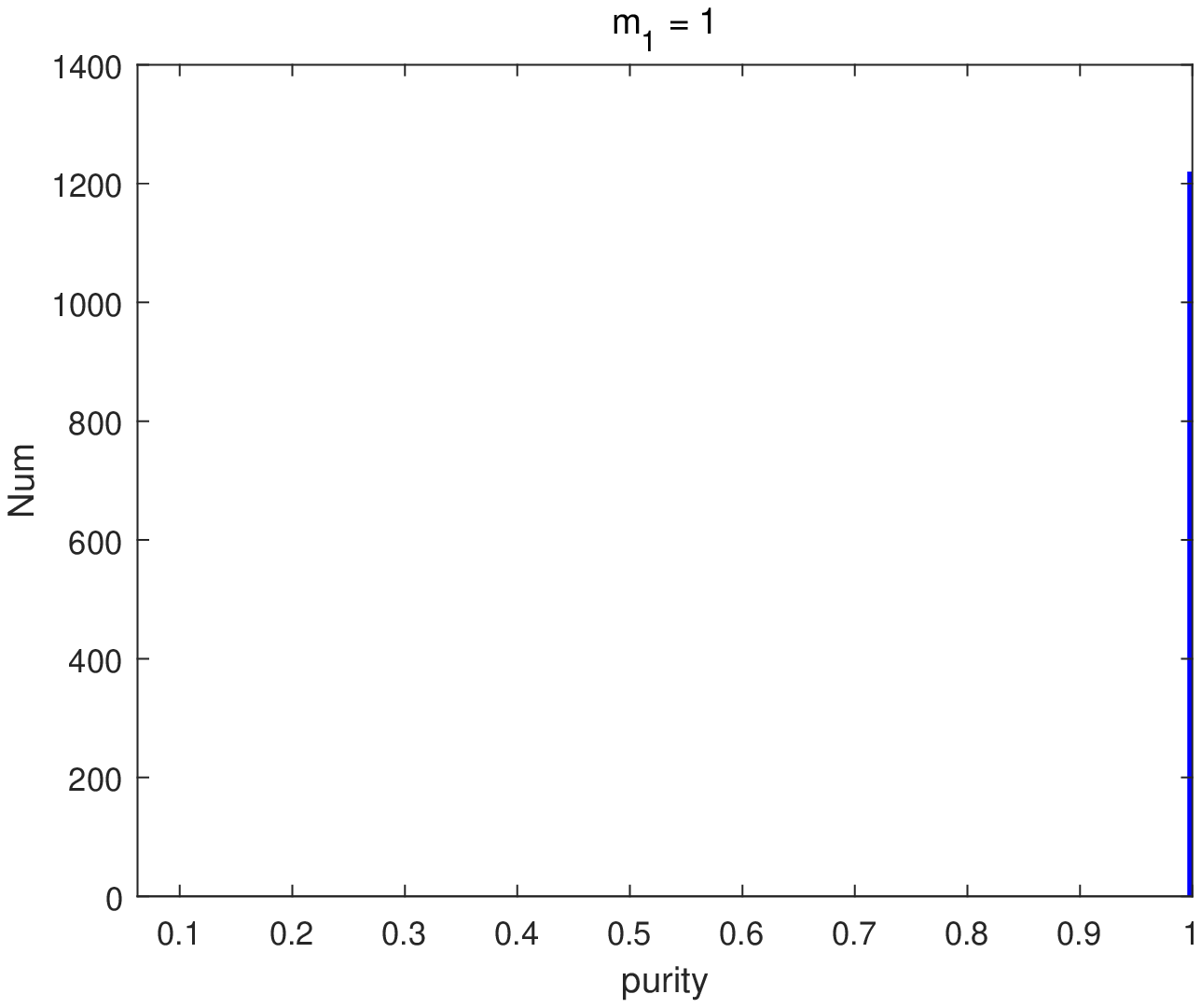}
                        \includegraphics[width=2in]{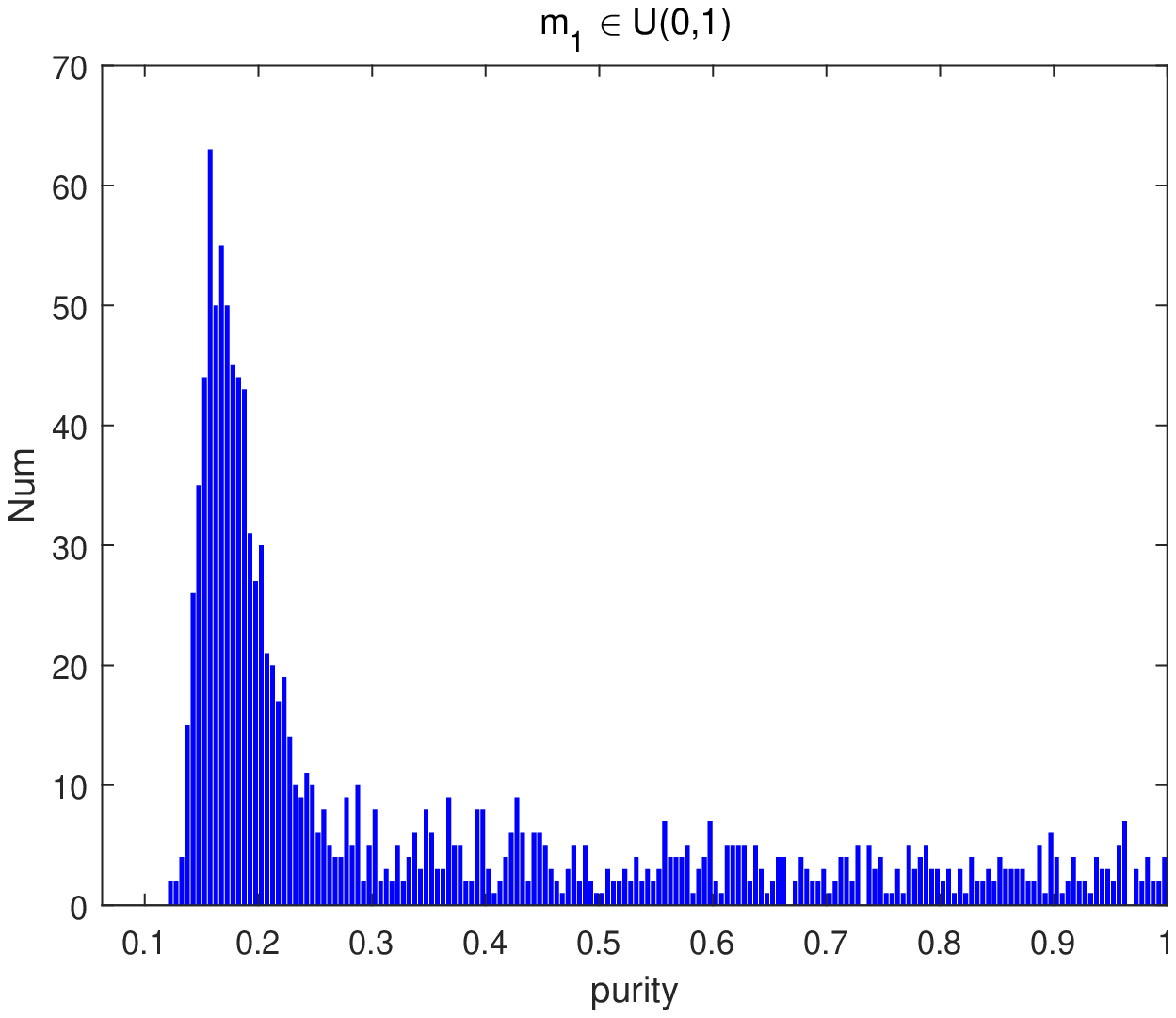}}
  \subfigure[$f_0$=0.8]{\includegraphics[width=2in]{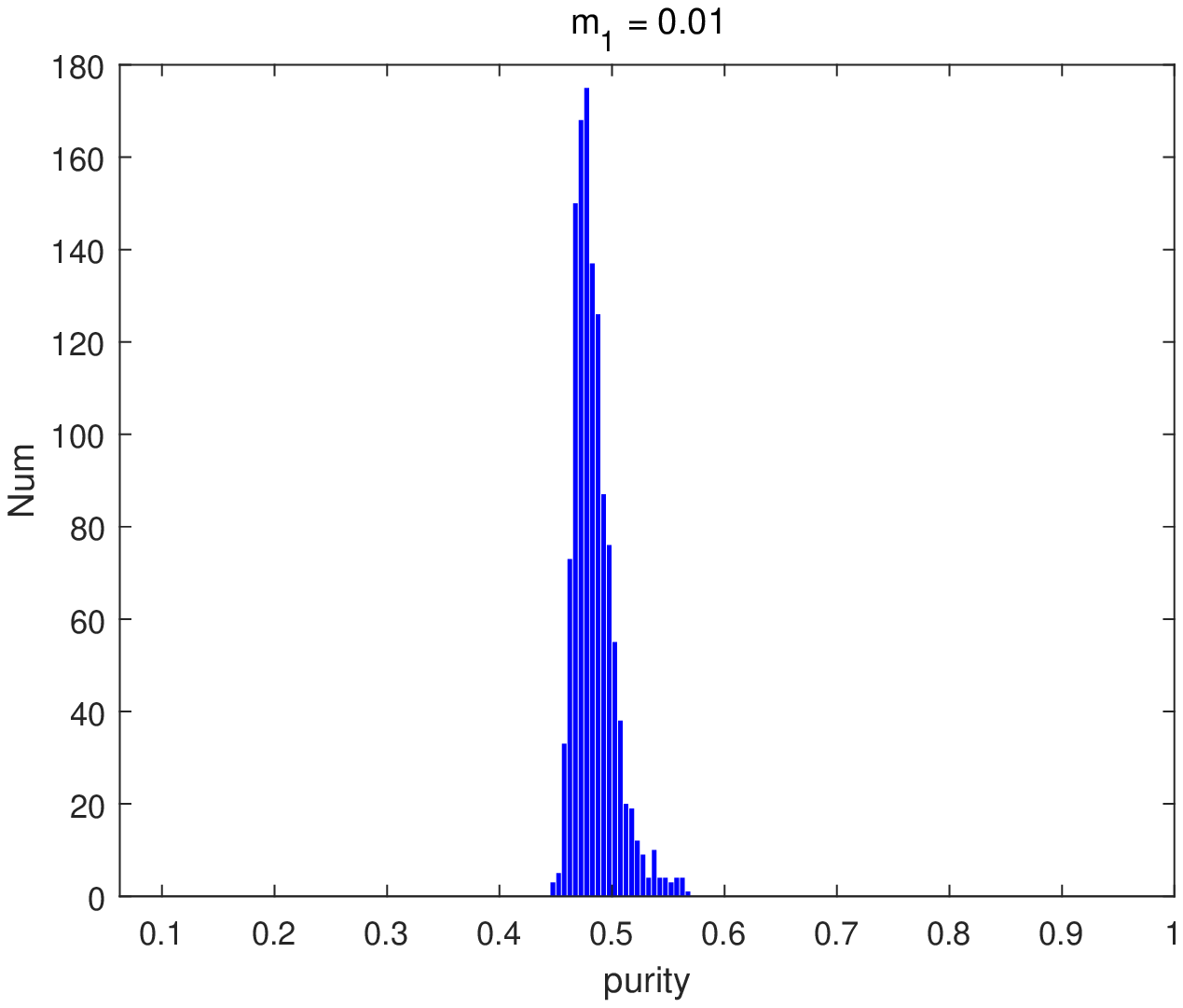}
               \includegraphics[width=2in]{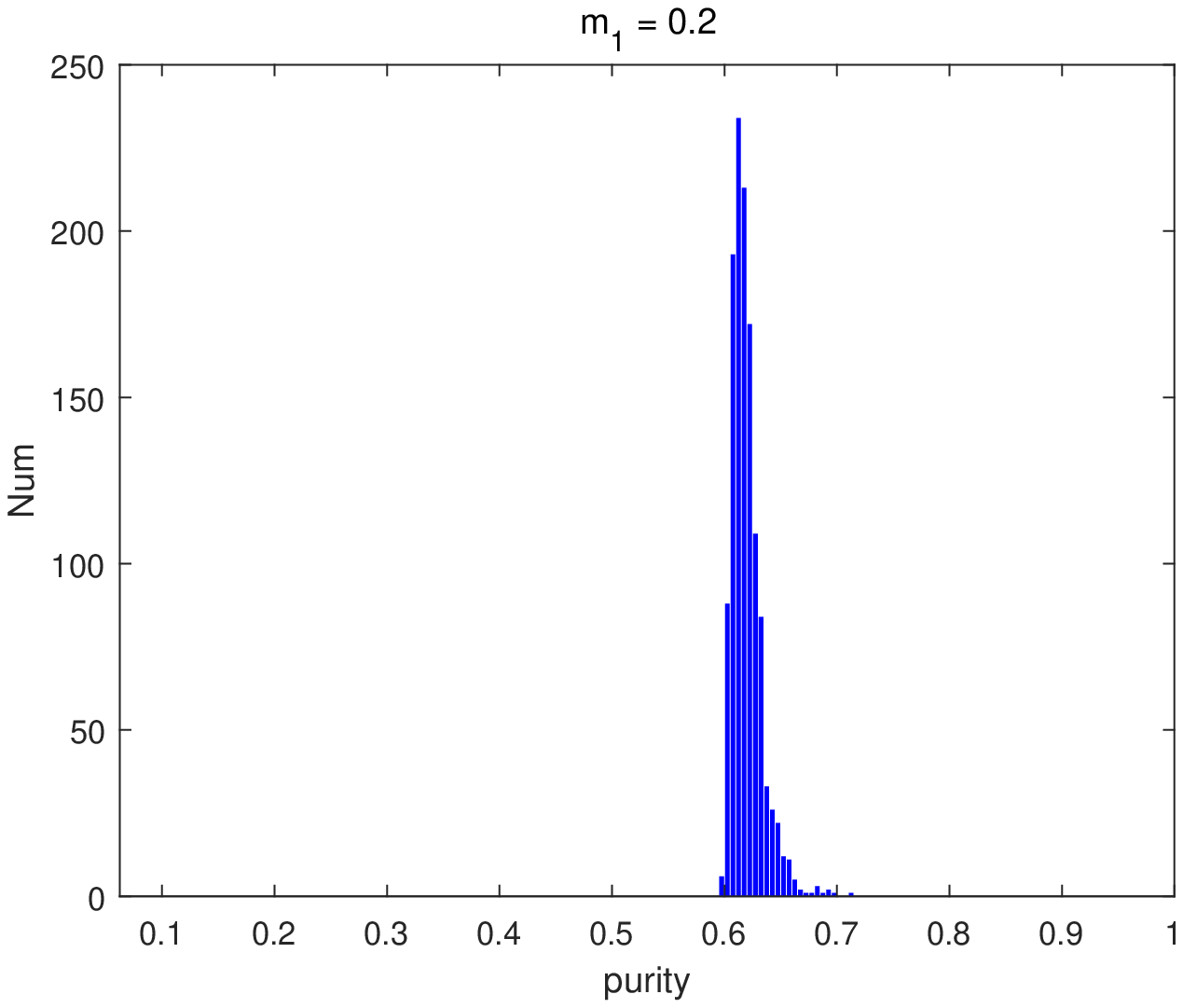}
               \includegraphics[width=2in]{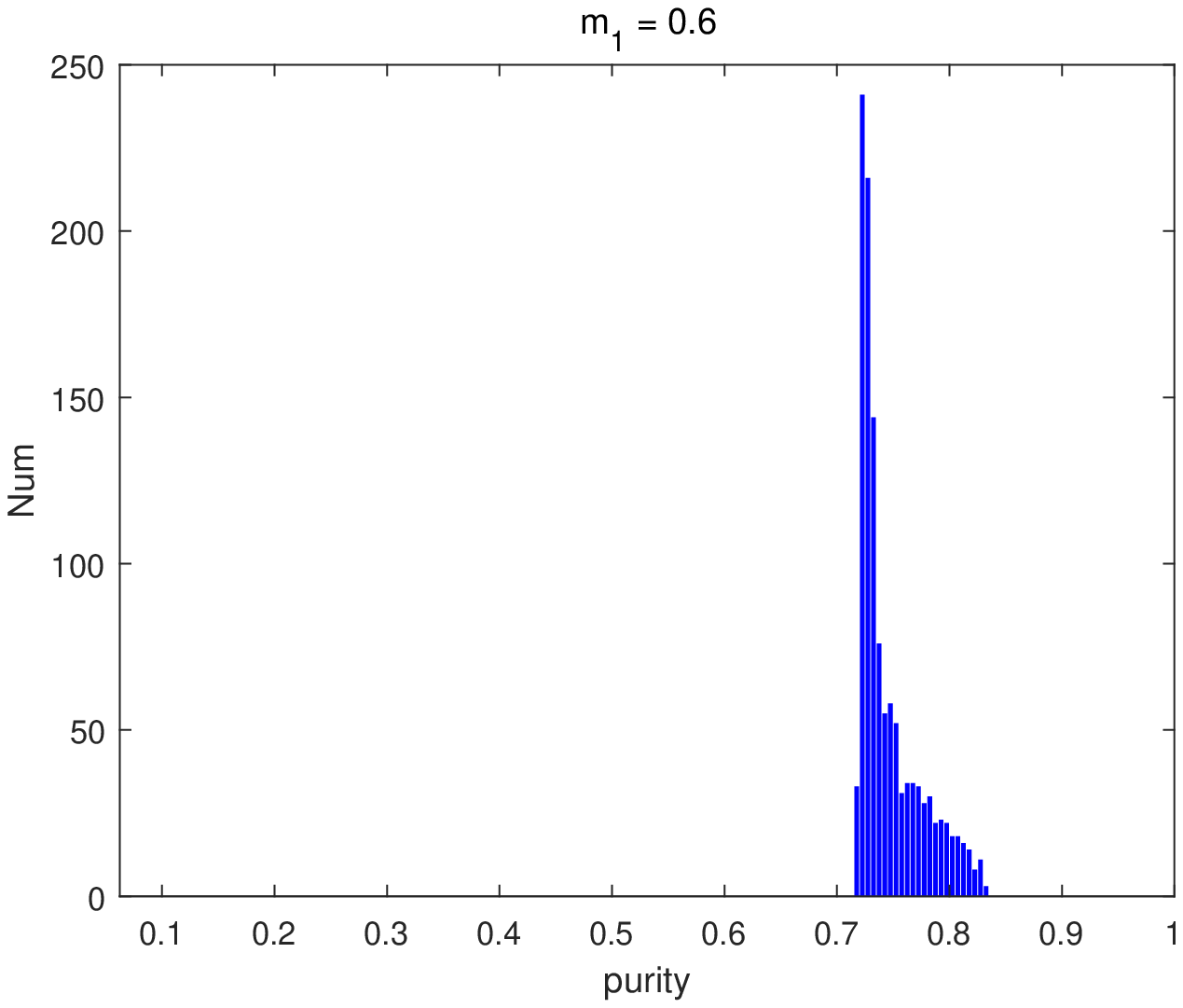}}
  \subfigure[$f_0$=0.8]{\includegraphics[width=2in]{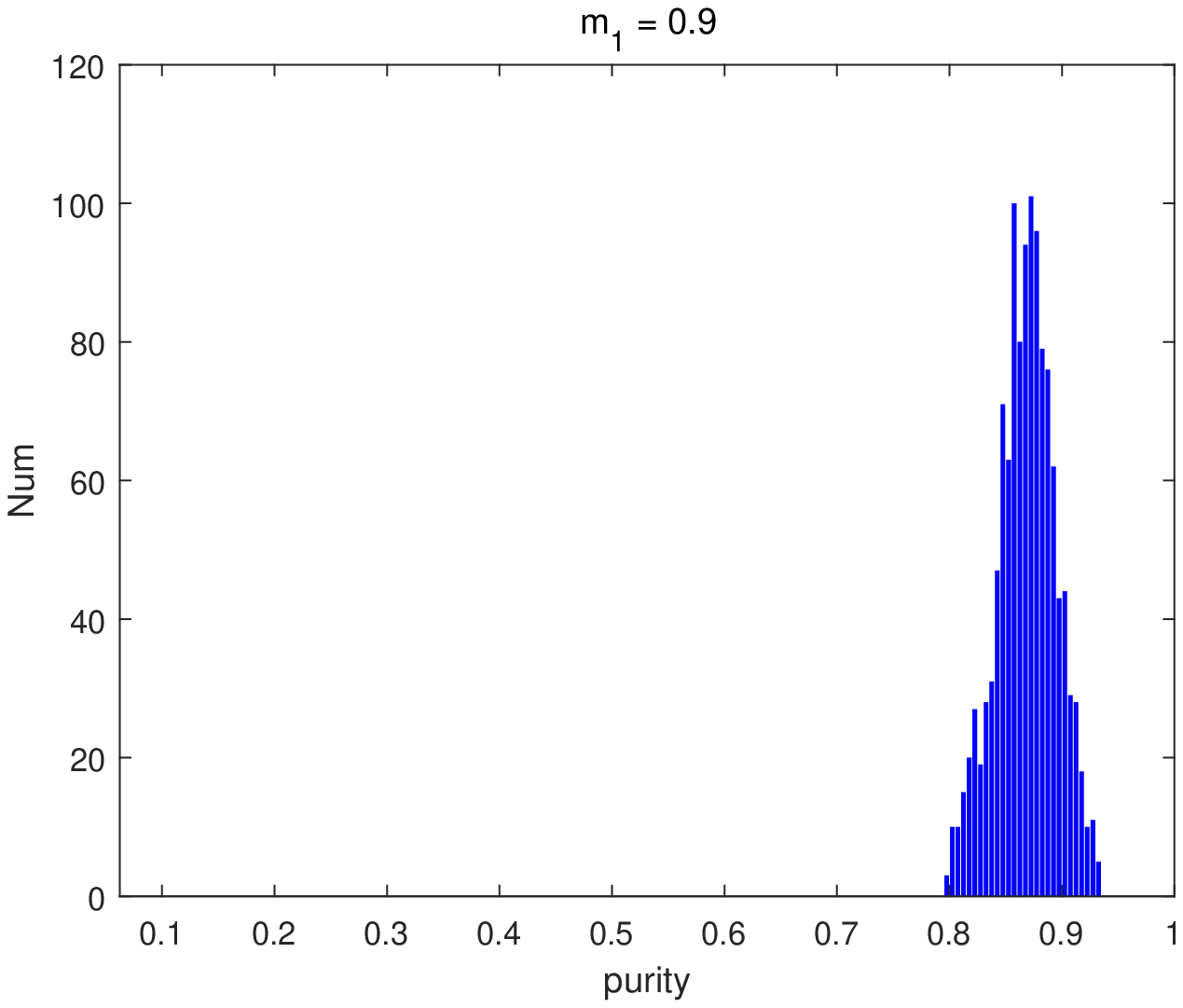}
                        \includegraphics[width=2in]{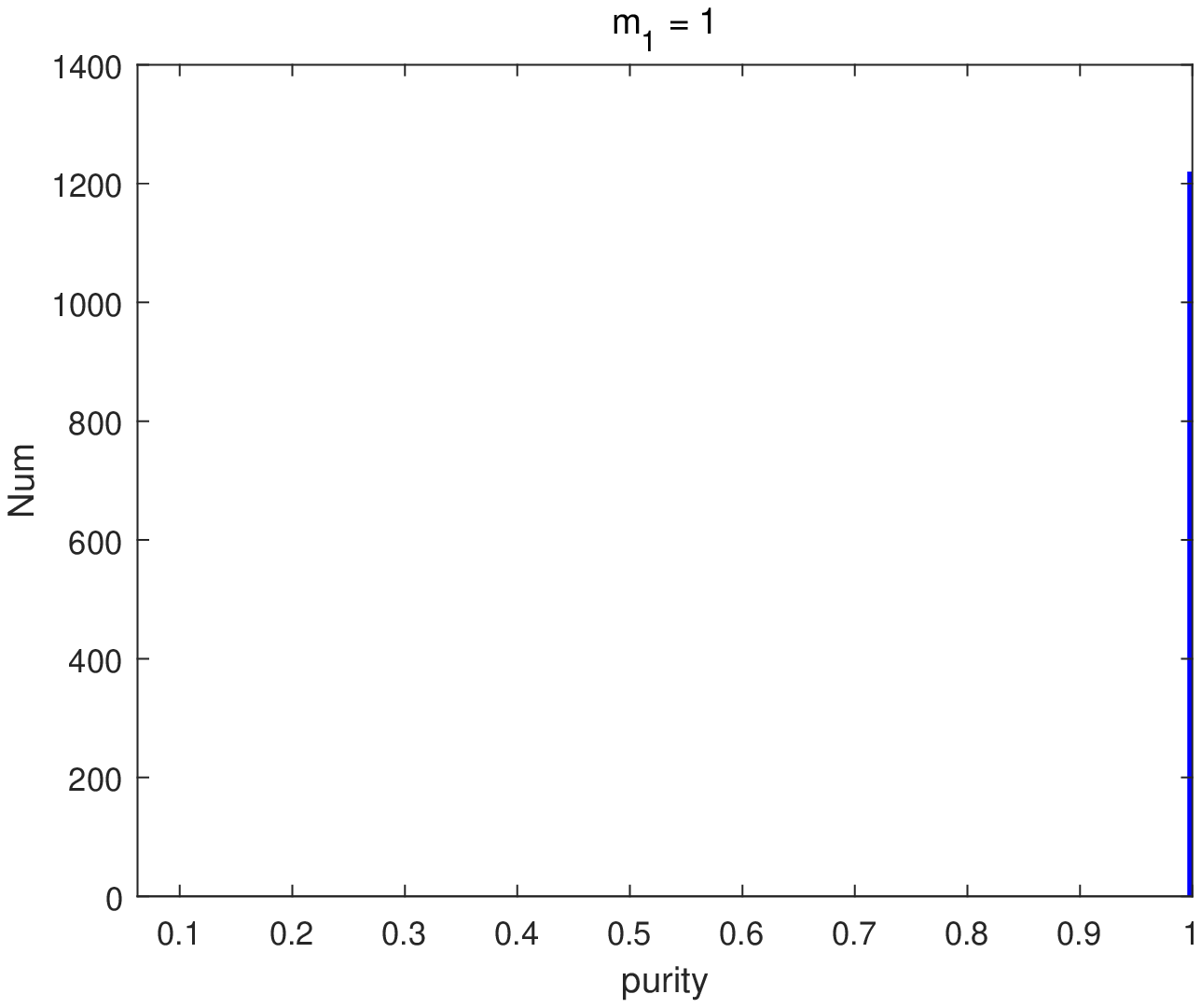}
                        \includegraphics[width=2in]{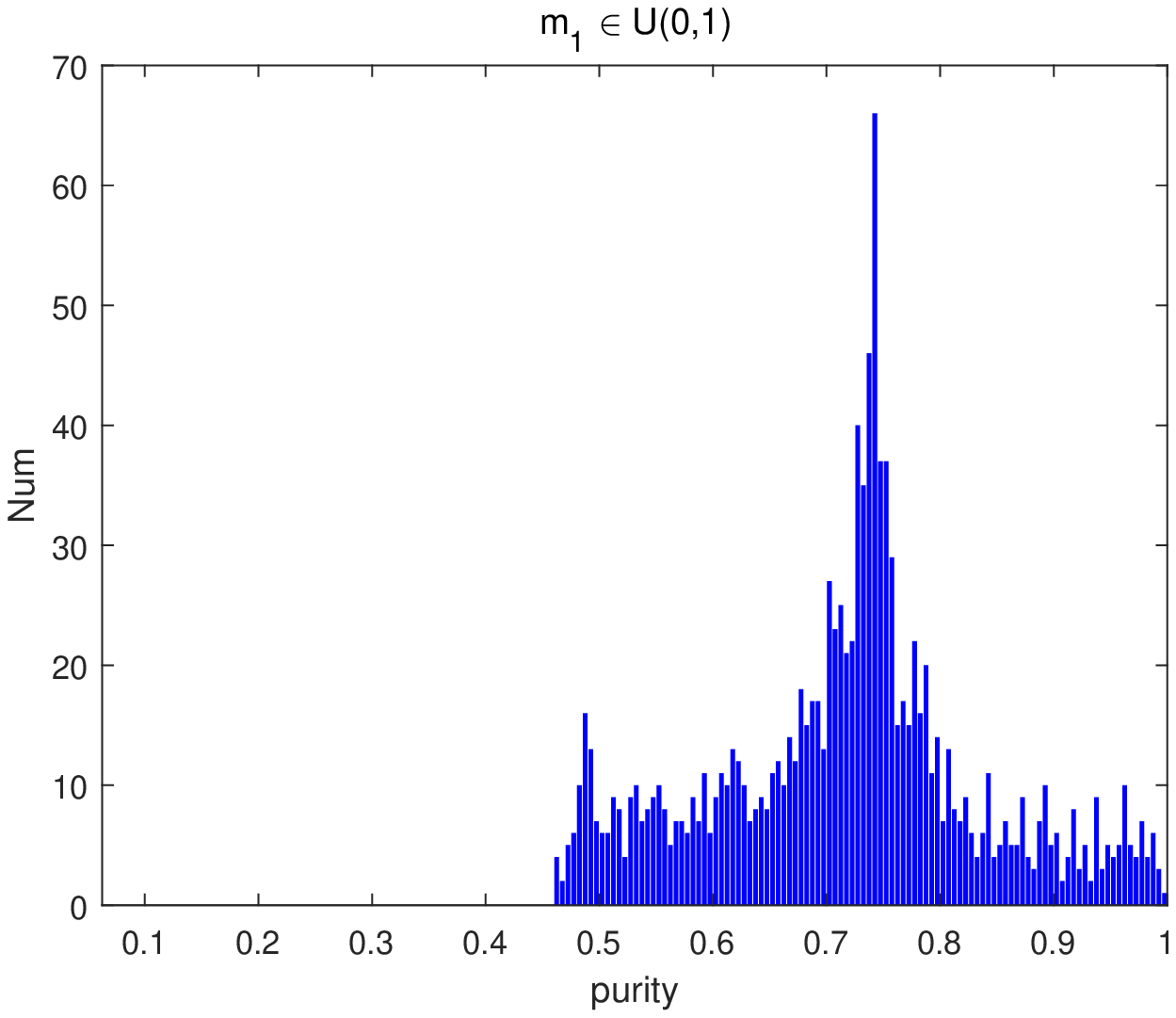}}
  \caption{(Color Online) Comparison of mixed-state distributions for different purity, where $m_1=1, 0.9, 0.6, 0.2, 0.01, U(0, 1)$. The specified quantum states fidelities are 0.25 and 0.8 respectively.}
\end{figure*}

Here, we discuss nine datasets in Table S3 for the four-qubit general state $|\varphi_4\rangle$, in which eight datasets belong to mixed states and the rest one to pure states. In eight mixed datasets, the purity distribution of $m_1$ is changed for seven datasets and the remaining one does not change the distribution of $m_1$.
\begin{table*}[!ht]
  \begin{center}
  \begin{tabular}{|c|c|c|}
  \multicolumn{3}{c}{\rule[-3mm]{0mm}{5mm} Table S3. Nine distributions of $m_1$}\\
  \hline
  ANN models       & datasets      & $m_1$ distribution         \\
  \hline
  A                & 1	            & $m_1=1-rand*rand$          \\
  \hline
  B                & 2              & $m_1=1-\sqrt{rand}*rand$   \\
  \hline
  C                & 3	            & $m_1$ belongs to a uniform distribution $U(0,1)$                    \\
  \hline
  D                & 4	            & $m_1$ belongs to a random standard normal distribution $N(0,1)$     \\
  \hline
  E                & 5	            & Pure states     \\
  \hline
  F                & 6	            & $m_1=1-rand*rand*rand$         \\
  \hline
  G                & 7	            & $m_1=1-rand*rand*rand*rand$    \\
  \hline
  H                & 8	            & $m_1=1-rand^3$         \\
  \hline
  I                & 9	            & $m_1=1-rand*rand^3$    \\
  \hline
  \end{tabular}
  \end{center}
\end{table*}

By comparing the distribution from Table S4, it can be seen that the distribution $m_1=1-rand^3$ is better than others because this distribution is effective for mixed states and pure states. It is worth note that the states in this distribution $m_1=1-rand^3$ are closer to pure states than those in other distributions. Here we only show the validation results of $k=3, 4, 5$ of nine ANN models for nine datasets because from these results can draw the conclusion we give.
\begin{table*}[!htbp]
  \begin{center}
  \begin{tabular}{|c|c|c|c|c|c|c|c|c|c|}
  \multicolumn{10}{c}{\rule[-3mm]{0mm}{5mm} Table S4. Accuracy of nine ANN models for nine datasets}\\
  \hline
  datasets   & 1       & 2         & 3      & 4        & 5         & 6        & 7       & 8          &9     \\
  \hline
  \diagbox{$k$}{model}   & \multicolumn{9}{c|}{A}    \\
  \hline
  3           & 79.78\%	&80.57\%   &81.11\%	 &77.19\%	&72.60\%	&78.09\%   &76.41\%	 &78.32\%	 &77.12\%      \\
  \hline
  4           & 86.99\%	&87.33\%   &87.42\%	 &83.37\%	&82.62\%	&86.34\%   &85.45\%	 &86.10\%	 &85.63\%    \\
  \hline
  5           & 89.82\%	&90.26\%   &89.94\%	 &85.96\%	&87.26\%	&90.01\%   &89.48\%	 &89.43\%	 &89.59\%       \\
  \hline
    \diagbox{$k$}{model}    & \multicolumn{9}{c|}{B}    \\
  \hline
  3           & 78.84\%	&80.24\%   &81.21\%	 &77.61\%	&67.65\%	&76.47\%	&73.82\%	&77.04\%	&74.98\%\\
  \hline
  4           & 86.34\%	&87.03\%   &87.36\%	 &83.75\%	&78.13\%	&84.70\%	&82.97\%	&84.84\%	&83.57\%\\
  \hline
  5           & 89.58\%	&90\%	   &89.70\%	 &86.49\%	&83.56\%	&88.89\%	&87.93\%	&88.65\%	&88.17\%\\
  \hline
  \diagbox{$k$}{model}    & \multicolumn{9}{c|}{C}    \\
  \hline
  3           & 77.53\%	&79.07\%	&80.29\%	&77.44\%	&66.83\%	&74.62\%	&71.98\%	&75.59\%	&75.59\%   \\
  \hline
  4           & 85.36\%	&86.32\%	&87.11\%	&83.61\%	&76.39\%	&83.14\%	&81.01\%	&83.69\%	&83.69\%    \\
  \hline
  5           & 88.67\%	&89.38\%	&89.72\%	&86.24\%	&81.36\%	&87.02\%	&85.34\%	&87.29\%	&87.29\%     \\
  \hline
  \diagbox{$k$}{model}    & \multicolumn{9}{c|}{D}    \\
  \hline
  3           & 75.29\%	 & 77.62\%	 & 79.37\%	 & 76.67\%	 & 59.93\%	 & 70.81\%	 & 67.10\%	 & 72.51\%	 & 69.18\%     \\
  \hline
  4           & 82.65\%	 & 84.39\%	 & 85.60\%	 & 82.80\%	 & 69.73\%	 & 79.13\%	 & 76.01\%	 & 80.27\%	 & 77.61\%     \\
  \hline
  5           & 85.72\%	 & 86.91\%	 & 87.74\%	 & 85.26\%	 & 76.09\%	 & 83.25\%	 & 80.90\%	 & 83.89\%	 & 82.05\%    \\
  \hline
  \diagbox{$k$}{model}    & \multicolumn{9}{c|}{E}    \\
  \hline
  3           & 22.93\%	 & 15.73\%	 & 11.10\%	 & 10.96\%	 & 85.05\%	 & 38.62\%	 & 54.07\%	 & 34.86\%	 & 46.63\%     \\
  \hline
  4           & 26.66\%	 & 18.21\%	 & 12.64\%	 & 12.63\%	 & 91.83\%	 & 44.57\%	 & 61.47\%	 & 39.29\%	 & 52.52\%     \\
  \hline
  5           & 27.60\%	 & 18.59\%	 & 12.49\%	 & 12.24\%	 & 95.07\%	 & 46.41\%	 & 64.04\%	 & 40.60\%	 & 54.55\%      \\
  \hline
  \diagbox{$k$}{model}    & \multicolumn{9}{c|}{F}    \\
  \hline
  3           & 80.02\%	 & 80.13\%	 & 80.18\%	 & 76.15\%	 & 75.52\%	 & 79.66\%	 & 78.62\%	 & 79.19\%	 & 78.86\%     \\
  \hline
  4           & 87.47\%	 & 86.94\%	 & 86.49\%	 & 82.64\%	 & 85.10\%	 & 87.66\%	 & 87.48\%	 & 86.93\%	 & 87.10\%   \\
  \hline
  5           & 90.45\%	 & 90.24\%	 & 89.55\%	 & 85.58\%	 & 90.70\%	 & 91.27\%	 & 91.37\%	 & 90.46\%	 & 91.04\%     \\
  \hline
  \diagbox{$k$}{model}    & \multicolumn{9}{c|}{G}    \\
  \hline
  3           & 79.13\%	 & 78.84\%	 & 78.65\%	 & 74.30\%	 & 78.77\%	 & 79.58\%	 & 79.67\%	 & 78.96\%	 & 78.39\%     \\
  \hline
  4           & 86.63\%	 & 85.97\%	 & 85.35\%	 & 81.16\%	 & 86.86\%	 & 87.63\%	 & 88.16\%	 & 86.78\%	 & 87.17\%    \\
  \hline
  5           & 90.43\%	 & 89.57\%	 & 88.58\%	 & 84.44\%	 & 91.87\%	 & 91.67\%	 & 92.25\%	 & 90.50\%	 & 90.92\%       \\
  \hline
  \diagbox{$k$}{model}    & \multicolumn{9}{c|}{H}    \\
  \hline
  3           & 79.09\%	 & 79.51\%	 & 80.78\%	 & 76.93\%	 & 76.03\%	 & 78.37\%	 & 77.45\%	 & 78.65\%	 & 78\%       \\
  \hline
  4           & 87.26\%	 & 87.15\%	 & 86.80\%	 & 83.11\%	 & 84.64\%	 & 87.09\%	 & 86.27\%	 & 86.75\%	 & 86.58\%    \\
  \hline
  5           & 90.65\%	 & 90.32\%	 & 89.24\%	 & 85.77\%	 & 89.30\%	 & 90.94\%	 & 90.38\%	 & 90.28\%	 & 90.59\%      \\
  \hline
  \diagbox{$k$}{model}     & \multicolumn{9}{c|}{I}    \\
  \hline
  3           & 79.73\%	 & 79.88\%	 & 79.15\%	 & 75.84\%	 & 76.82\%	 & 79.62\%	 & 79.18\%	 & 79.33\%	 & 78.65\%     \\
  \hline
  4           & 86.82\%	 & 86.36\%	 & 86.39\%	 & 82.18\%	 & 86.92\%	 & 87.30\%	 & 87.53\%	 & 86.79\%	 & 87.39\%    \\
  \hline
  5           & 90.08\%	 & 89.46\%	 & 88.81\%	 & 85.18\%	 & 91.23\%	 & 91\%	     & 91.46\%	 & 90.26\%	 & 91.08\%   \\
  \hline
  \end{tabular}
  \end{center}
\end{table*}

\vspace{4mm}
\centerline{\textbf{F. Universality of our method}}
We will prove the universality of our methods for preparing quantum states. To better demonstrate this inference, we show three dataset classes with three six-qubit general target pure states $|\varphi_6^1\rangle, |\varphi_6^2\rangle, |\varphi_6^3\rangle$, respectively. By analyzing the results from Table S5, we conclude that our method of preparing database is universal, corresponding to an arbitrary target pure state.

\begin{table*}[!htbp]
  \begin{center}
  \begin{tabular}{|p{2cm}|p{2cm}|p{2cm}|p{2cm}|}
  \multicolumn{4}{c}{\rule[-3mm]{0mm}{5mm} Table S5. Datasets comparison of three six-qubit general states}\\
  \hline
  \multicolumn{4}{|c|}{six-qubit (Hid-neu:700-300, precision with $\pm 1\%$)}                                     \\
  \hline
  \diagbox{$k$}{states}        & $|\varphi_6^1\rangle$      & $|\varphi_6^2\rangle$      & $|\varphi_6^3\rangle$       \\
  \hline
  2                            & 79.17\%	                &77.76\%	                 &78.77\%\\
  \hline
  3                            & 86.89\%	                &85.39\%	                 &88.03\%\\
  \hline
  4                            & 91.52\%	                &90.56\%	                 &92.49\%\\
  \hline
  5                            & 94.09\%	                &93.68\%	                 &94.48\%\\
  \hline
  6                            & 95.21\%	                &95.19\%	                 &94.81\%\\
  \hline
  7                            & 97.10\%	                &96.24\%	                 &97.15\%\\
  \hline
  \end{tabular}
  \end{center}
\end{table*}
\vspace{4mm}
\centerline{\textbf{G. Poisson noise analysis}}
In an experiment, noise is inevitable. In our work, we assume that the noise follows the Poisson law. Using the \emph{random} function in Matlab, we can obtain one value from the Poisson distribution with $NP$, in which $P$ is the basis measurements, and $NP$ is the ideal coincidence count.

We give a comparison of different number of samples for a four-qubit general state (See Fig.5). The noise model uses the number of samples of $N=$1,000, 4000, 7,000, 10,000, 10,0000 and 1,000,000. For comparison, we find that the noise model for the number of sample $N=$10,000 is appropriate. Therefore, these datasets are generated with the number of samples $N=$10,000 from two-qubit to seven-qubit states.

\begin{figure*}[!htbp]
  \centering
  \includegraphics[width=0.5\textwidth]{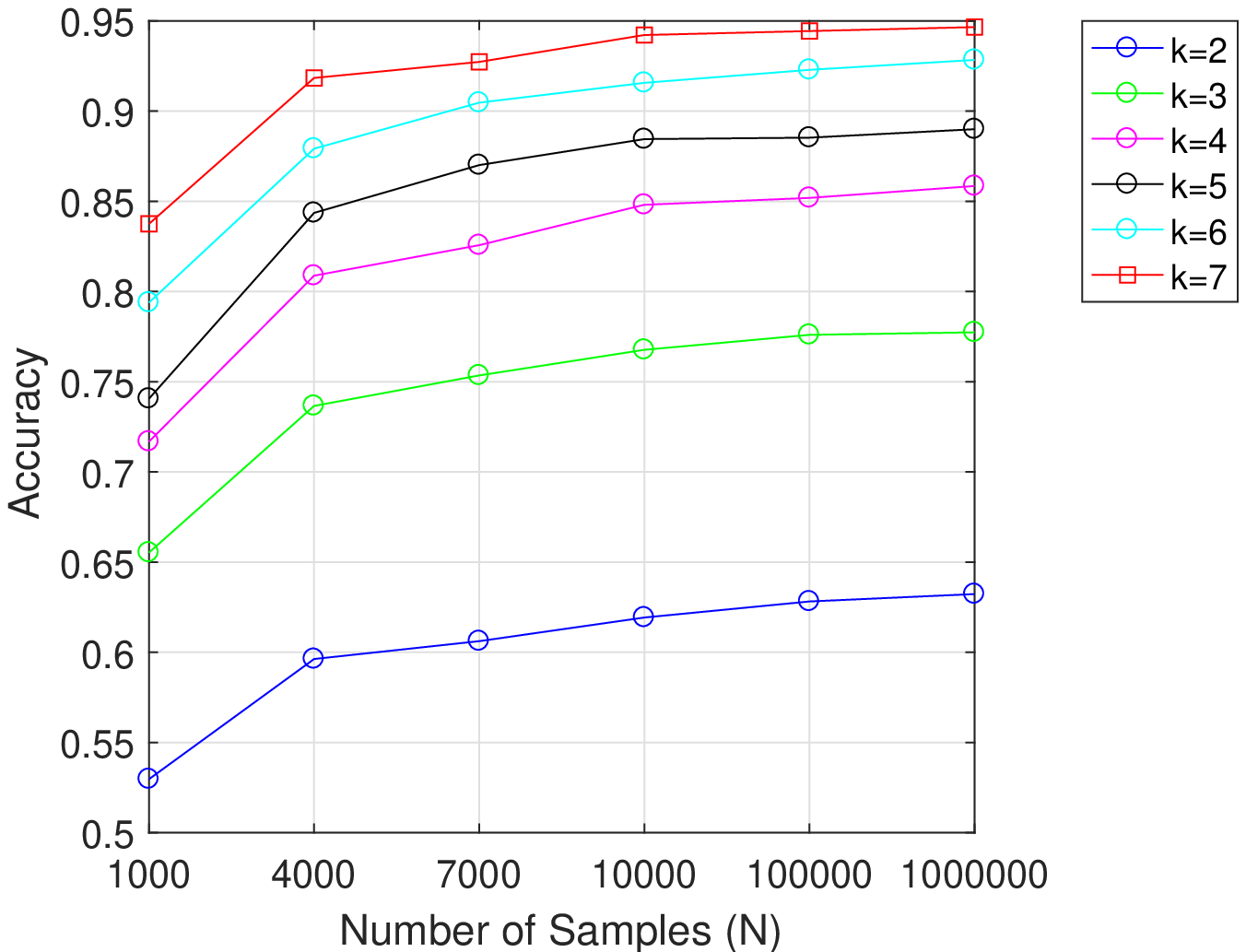}
  \caption{(Color online) Comparison of the accuracy of neural network models with different number of samples. Taking a four-qubit general state as an example, when $k=2,3,4,5,6,7$, the neural network models with the number of samples 1000, 4000, 7000, 10000, 100000, 100000 are generated respectively. The higher the number of Pauli operator measurement settings used, the higher the prediction accuracy is. The higher of the number of samples $N$ is, the higher the accuracy is.}\label{A1}
\end{figure*}

\section{Selection of Pauli combinations}
Let $W_k$ $(k=1,2,...,2^{2n})$ denotes all possible Pauli operators ($n$-fold tensor products of $I$, $X$, $Y$ and $Z$). Then $tr(\rho\sigma)$ can be rewritten in Ref.\cite{40Miszczak} as follows.
\begin{eqnarray}
tr(\rho\sigma)=\sum_k\frac{1}{d}\chi_\rho(k)\chi_\sigma(k),
\end{eqnarray}
where the characteristic function is defined as $\chi_\rho(k)=tr(\rho W_k)$. For a target pure state $\rho$, we define the weight $Pr(k)$ corresponding to the Pauli operators  $W_k$.
\begin{eqnarray}
Pr(k)=[\chi_\rho(k)]^2,
\end{eqnarray}
If we want to know the fidelity between a target pure state and an arbitrary state, we need to get $\chi_\sigma(k)$ in Eq.(23) related to the weight $Pr(k)\neq0$. Obviously, the general state requires more Pauli combinations than a special state. Here we have chosen the $k$ Pauli operators with the largest absolute value of the expectation value of the desired quantum state.

Here we show the accuracy of the neural network for the Pauli operators with a large absolute value of expectation and for the Pauli operators with a small absolute value of expectation, respectively (See Table S6). Take the four-qubit cluster state and the general state as an example, the ANN model for the Pauli operator with a large absolute value of expectation can get more information than that for the Pauli operator with a small absolute value of expectation.
\begin{table*}[!htbp]
  \begin{center}
  \begin{tabular}{|c|c|c|c|c|}
  \multicolumn{5}{c}{\rule[-3mm]{0mm}{5mm} Table S6. Comparison of two types of weights}\\
  \hline
  \multicolumn{5}{|c|}{four-qubit states (Precision $\pm 1\%$)}                                                \\
  \hline
  \diagbox{$k$}{states}    & $|\varphi_4\rangle$(largest weight,         &  $|\varphi_4\rangle$(smallest weight,        & Cluster (weight=1,        & Cluster (weight=0,        \\
                               & Hid-neu:5000)                             &  Hid-neu:5000)                            & Hid-neu:3000)             & Hid-neu:3000)         \\
  \hline
  2                            & 61.48\%	                       &56.35\%	                              &90.75\%	                     &67.99\%     \\
  \hline
  3                            & 77.18\%	                       &69.43\%	                              &96.45\%	                     &81.98\%      \\
  \hline
  4                            & 84.51\%	                       &77.79\%	                              &98.48\%	                     &91\%   \\
  \hline
  5                            & 88.47\%	                       &83.70\%	                              &99.37\%	                     &96.18\%      \\
  \hline
  6                            & 91.80\%	                       &90.76\%                               &99.81\%	                     &97.34\%      \\
  \hline
  7                            & 94.21\%	                       &92.83\%	                              &99.96\%	                     &97.67\%   \\
  \hline
  \end{tabular}
  \end{center}
\end{table*}

We then show the selected Pauli operators from two-qubit to eight-qubit states in Table S7.
\begin{table*}[!htbp]
  \begin{center}
  \begin{tabular}{|c|c|c|c|}
  \multicolumn{4}{c}{\rule[-3mm]{0mm}{5mm} Table S7. Pauli operators}\\
  \hline
  \multicolumn{2}{|c|}{two-qubit}                                                      &  \multicolumn{2}{c|}{three-qubit} \\
  \hline
  states               & Pauli operators                   & states             & Pauli operators\\
  \hline
  Bell                &XX;YZ;ZY;YY;ZX;XZ;XY                & GHZ                & ZZZ;XXX;XYY;YXY;YYX; YYY;XXZ\\
  $W_2$               &XX;YZ;ZY;XZ;YY;ZX;ZZ                & W                  & ZZZ;ZXX;ZYY;XZX;XXZ;YZY;YYZ\\
  $|\varphi_2\rangle$ &XY;YX;ZZ;YZ;XX;ZY;YY;ZX;XZ          & $|\varphi_3\rangle$& YYY;XXY;ZZY;XYZ;YXX;YXZ;YXY \\
  \hline
   \multicolumn{4}{|c|}{four-qubit}                                                   \\
  \hline
  Cluster             & \multicolumn{3}{c|}{ZZXX;ZZYY;XXZZ;YYZZ;XYXY;XYYX;YXXY} \\
  Dicke               & \multicolumn{3}{c|}{XXXX;YYYY;ZZZZ;XXZZ;ZZYY;ZZXX;YYZZ} \\
  GHZ                 & \multicolumn{3}{c|}{XXXX;YYYY;ZZZZ;XXYY;XYXY;YXYX;YYXX} \\
  W                   & \multicolumn{3}{c|}{ZZZZ;ZZXX;ZZYY;XXZZ;YYZZ;XZXZ;YZYZ} \\
  $|\varphi_4\rangle$ & \multicolumn{3}{c|}{YXXZ;XYZX;ZZYY;YZXX;XYZY;YXZZ;ZZYX} \\
  \hline
  \multicolumn{4}{|c|}{five-qubit}                                                \\
  \hline
  Cluster             & \multicolumn{3}{c|}{XZZXZ;ZYXYZ;ZXZZX;YXXXY;YYZZX;XZZYY;ZYXXY}  \\
  C-ring              & \multicolumn{3}{c|}{XXXXX;ZYXYZ;ZZYXY;XYZZY;YXYZZ;YZZYX;XZZYX}  \\
  Dicke               & \multicolumn{3}{c|}{ZZZZZ;XXXXZ;YYYZY;ZZZXX;ZYYYY;YZZYZ;XXXZX} \\
  GHZ                 & \multicolumn{3}{c|}{XXXXX;ZZZZZ;XYXXY;XYXYX;XYYXX;YXYYY;YYXYY} \\
  W                   & \multicolumn{3}{c|}{ZZZZZ;XXZZZ;XZZXZ;YYZZZ;YZZYZ;ZXZXZ;ZYZYZ} \\
  $|\varphi_5\rangle$ & \multicolumn{3}{c|}{YYXZX;XZZYZ;ZYYYY;XZYXX;XXZZY;YZXXZ;ZYXZY} \\
  \hline
  \multicolumn{4}{|c|}{six-qubit}                                                \\
  \hline
  C23                 & \multicolumn{3}{c|}{XZXYXY;XZXYXY;ZXYZYZ;ZYZYXZ;YXYXZX;XZXZYY;ZYZZXY}  \\
  Dicke               & \multicolumn{3}{c|}{ZZZZZZ;XXZZZZ;ZZZZYY;ZZYYZZ;ZZXZZX;ZZZXXZ;YYZZZZ}  \\
  GHZ                 & \multicolumn{3}{c|}{XXXXXX;YYYYYY;ZZZZZZ;XXXYYY;XYYYYX;YXYYXY;YYYYXX} \\
  W                   & \multicolumn{3}{c|}{ZZZZZZ;XZXZZZ;XZZZZX;YYZZZZ;YZZZYZ;ZXZXZZ;ZYZZYZ} \\
  $|\varphi_6\rangle$ & \multicolumn{3}{c|}{XXYXYX;ZZXYZZ;YYZYXZ;ZZXZXY;ZXYXYY;ZYZZZX;YZZYZY} \\
  \hline
  \multicolumn{4}{|c|}{seven-qubit}\\
  \hline
  $|\varphi_7\rangle$ & \multicolumn{3}{c|}{XYZXXXX;ZZXZYYY;YXXYZZZ;ZYYZZZZ;XXZYYYX;YZXZXXY;ZYYXYZX} \\
  \hline
  \multicolumn{4}{|c|}{eight-qubit}\\
  \hline
  $|\varphi_8\rangle$ & \multicolumn{3}{c|}{XXXXXXXX;YYYYYYYY;ZZZZZZZZ;XYZXYZXY;YZXYZXYZ} \\
  \hline
  \end{tabular}
  \end{center}
\end{table*}

\section{Precision of fidelity estimation}
Figure 6 presents some important information about the neural network with different number of labels. In Fig.6(a-c), in the case of a fixed number of labels, the higher the number of Pauli operator measurement settings used, the higher the prediction accuracy is. Meanwhile, the higher the fidelity of the predicted quantum states is, the higher the accuracy is. In Fig.6(d), the more quantum state fidelity intervals are divided, the higher the accuracy is. However, the accuracy of the neural network model with 234 labels is not much higher than that of the neural network model with 122 labels. Moreover, the higher the number of labels are, the more resources and time are consumed, so the neural network with 122 labels is selected as the most appropriate. The fidelity interval of the quantum state is respectively divided into 66 labels, 122 labels and 234 labels for the specific interval in Ref.\cite{67three}.

\begin{figure*}[htbp]
  \flushleft
  \subfigure[\ the state $|\varphi_5\rangle$ with 66 labels]{\includegraphics[width=0.45\textwidth]{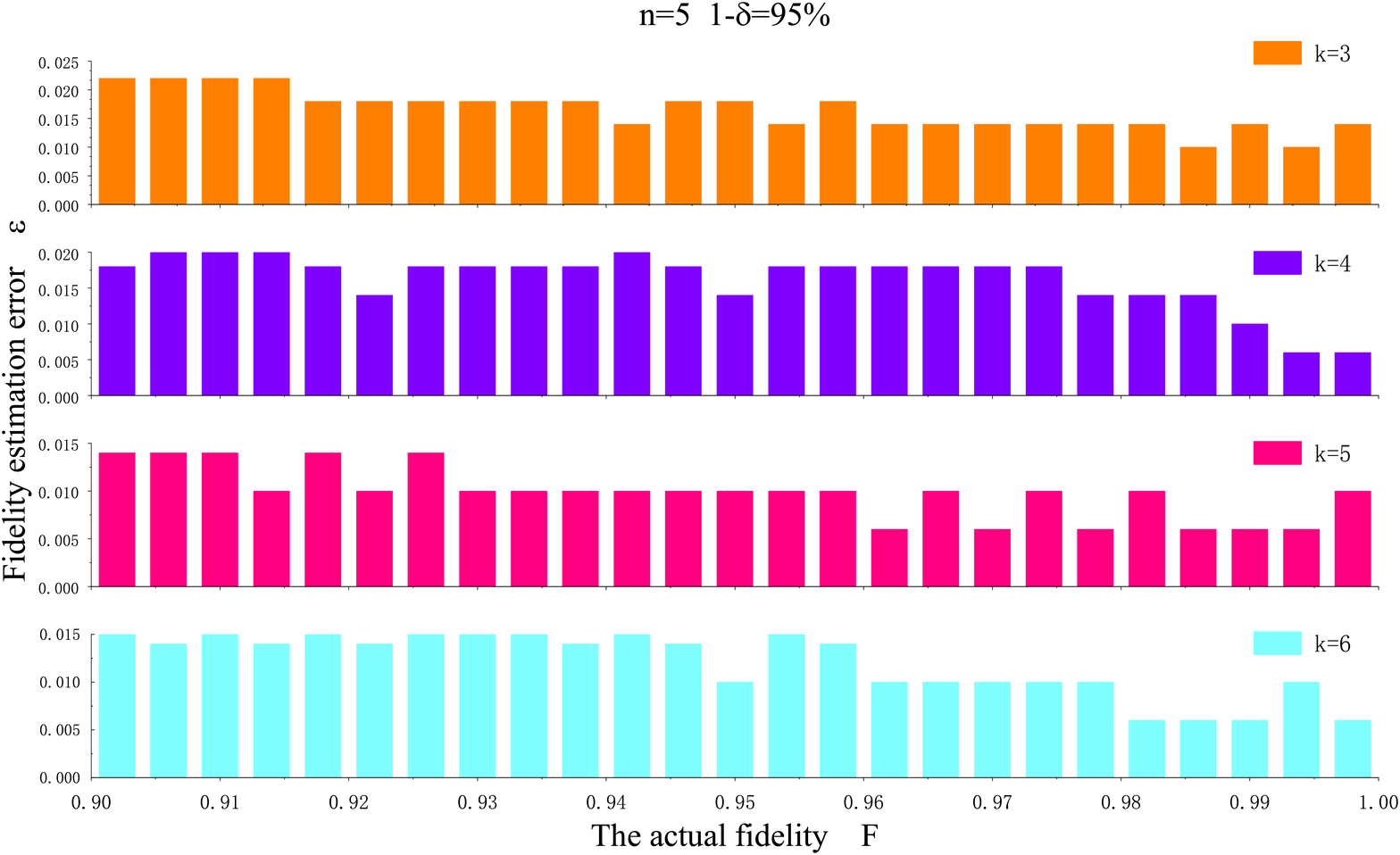}}\quad
  \subfigure[\ the state $|\varphi_5\rangle$ with 122 labels]{\includegraphics[width=0.45\textwidth]{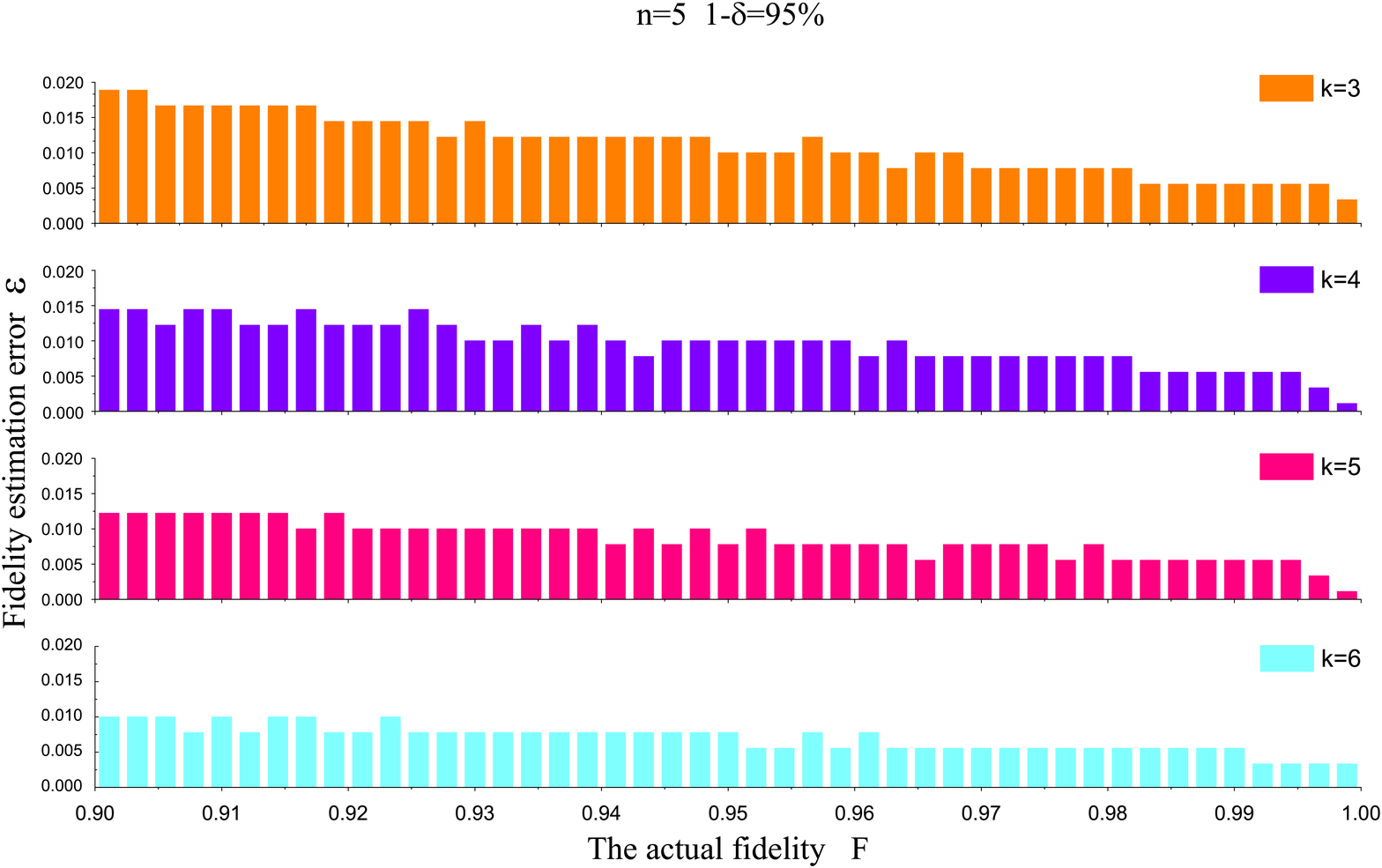}}

  \subfigure[\ the state $|\varphi_5\rangle$ with 234 labels]{\includegraphics[width=0.45\textwidth]{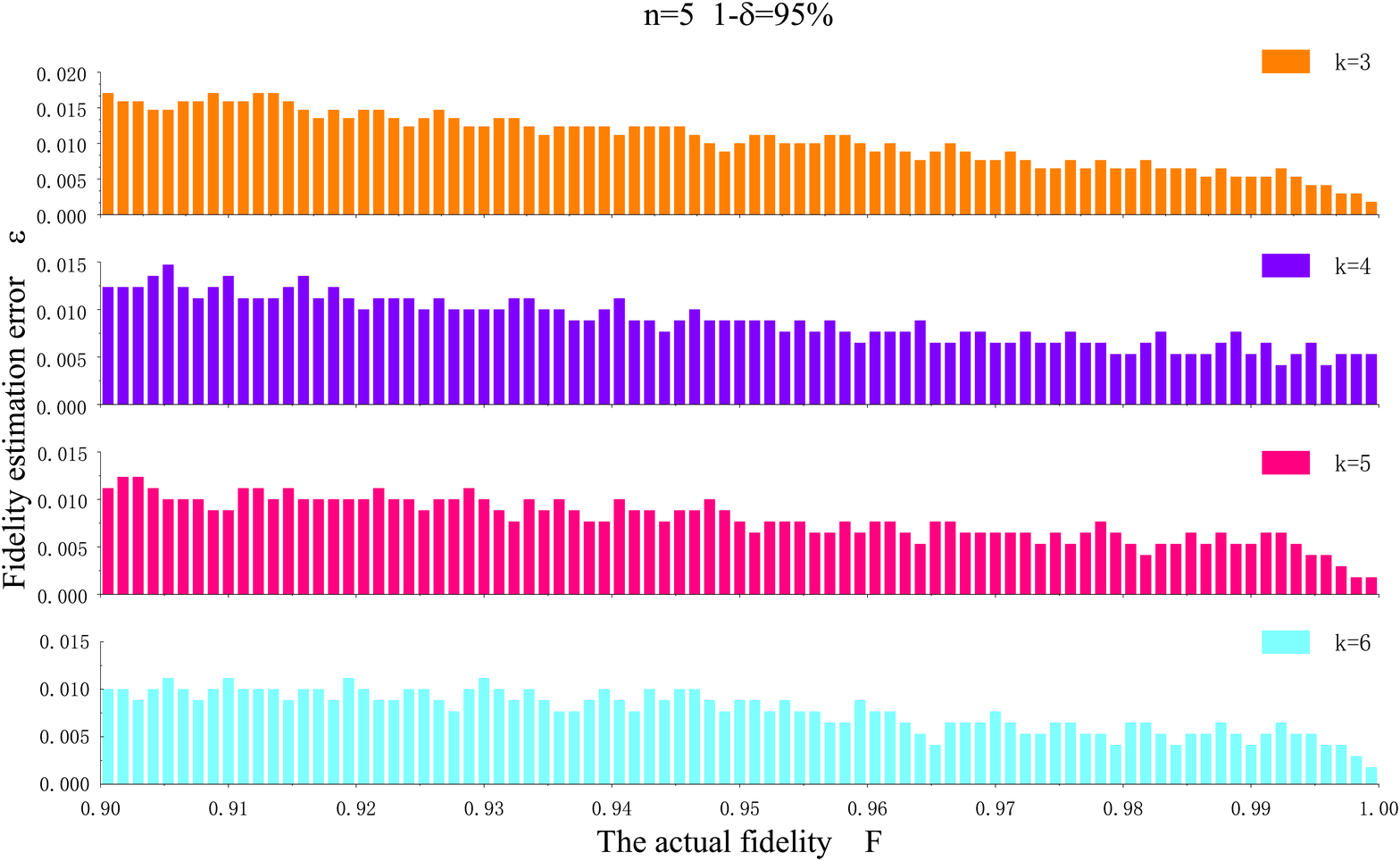}}
  \subfigure[\ the state $|\varphi_5\rangle$ with 66 labels, 122 labels and 234 labels]{\includegraphics[width=0.45\textwidth]{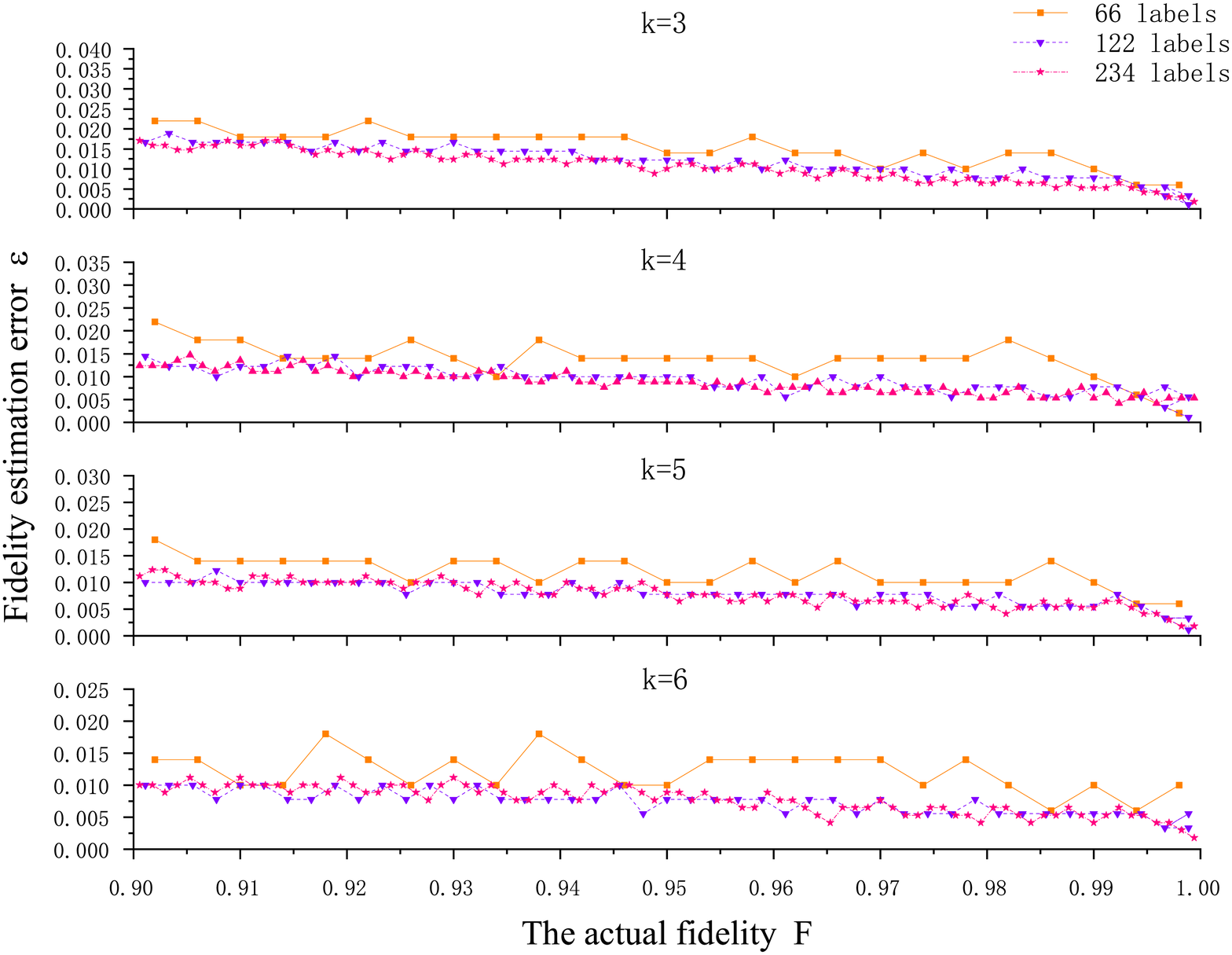}}
  \caption{(Color Online) A plot of the prediction accuracy of the neural network versus the quantum state fidelity when measurements are made using three, four, five, and six Pauli operator measurement settings. Here we choose a 5-qubit general state $|\varphi_5\rangle$ as the target state. The higher the number of Pauli operator measurement settings used, the higher the prediction accuracy is. The higher the fidelity of the predicted quantum states is, the higher the accuracy is. The more quantum state fidelity intervals are divided, the higher the accuracy is. However, the higher the number of labels are, the more resources and time are consumed, so the neural network model with 122 labels is selected as the most appropriate.}
\end{figure*}

\section{Accuracies of ANN models for N-qubit states}
In Table S8, we show all the accuracies of the ANN models from two-qubit to seven-qubit states with the precision $\pm 1\%$.
\begin{table*}[!htbp]
  \begin{center}
  \begin{tabular}{|c|c|c|c|c|c|c|}
  \multicolumn{7}{c}{\rule[-3mm]{0mm}{5mm} Table S8. Accuracies of ANN models}\\
  \hline
  \multirow{2}{*}{\diagbox{$k$}{states}}   & \multicolumn{3}{c}{two-qubit (Hid-neu:2000)}           & \multicolumn{3}{|c|}{three-qubit (Hid-neu:2000)}    \\
  \cline{2-7}
                               &Bell        &$W_2$      &$|\varphi_2\rangle$          &GHZ       &W        &$|\varphi_3\rangle$    \\
  \hline
  2                            & 30.81\%	&46.65\%	&36.56\%	                  &34.65\%	 &57.35\%  &54.22\%   \\
  \hline
  3                            & 50.50\%	&64.87\%	&62.50\%	                  &62.55\%	 &68.10\%  &68.38\%	 \\
  \hline
  4                            & 76.73\%	&78.68\%	&76.25\%	                  &70.40\%	 &76.78\%  &80.10\% \\
  \hline
  5                            & 90.16\%	&92.21\%	&80.66\%	                  &77.75\%	 &86.65\%  &89.49\%	   \\
  \hline
  6                            & 99.99\%	&93.90\%	&87.27\%	                  &81.23\%	 &89.07\%  &92.72\%   \\
  \hline
  7                            & 99.99\%	&99.99\%	&93.38\%	                  &87.73\%	 &89.87\%  &95.38\%    \\
  \hline
  \multirow{2}{*}{\diagbox{$k$}{states}}  & \multicolumn{6}{c|}{four-qubit (Hid-neu:2000)}            \\
  \cline{2-7}
                               &Cluster        &GHZ      &W          &Dicke           &\multicolumn{2}{|c|}{$|\varphi_4\rangle$}      \\
  \hline
  2                            &90.66\%	       &78.19\%	 &86.86\%	 &91.56\%	      &\multicolumn{2}{|c|}{61.93\%}  \\
  \hline
  3                            &96.47\%	       &97.82\%	 &89.25\%	 &96.77\%	      &\multicolumn{2}{|c|}{76.77\%} \\
  \hline
  4                            &98.51\%	       &98.53\%	 &93.58\%	 &99.41\%	      &\multicolumn{2}{|c|}{84.81\%} \\
  \hline
  5                            &99.40\%	       &99.34\%	 &95.53\%	 &99.80\%	      &\multicolumn{2}{|c|}{88.45\%} \\
  \hline
  6                            &99.79\%	       &99.74\%	 &96.83\%	 &99.97\%	      &\multicolumn{2}{|c|}{91.57\%}	  \\
  \hline
  7                            &99.96\%	       &99.94\%	 &97.92\%	 &99.99\%	          &\multicolumn{2}{|c|}{94.22\%}  \\
  \hline
  \multirow{2}{*}{\diagbox{$k$}{states}}   & \multicolumn{6}{|c|}{five-qubit (Hid-neu:500-300)}            \\
  \cline{2-7}
                               &Cluster        &C-ring      &Dicke          &GHZ          &W         &$|\varphi_5\rangle$    \\
  \hline
  2                            &87.26\%	       &79.87\%	    &84.34\%	    &92.28\%	  &81.67\%	 &70.40\% \\
  \hline
  3                            &89.73\%	       &82.96\%	    &89.08\%	    &93.15\%	  &91.49\%	 &81.75\% \\
  \hline
  4                            &90.48\%	       &84.52\%	    &92.62\%	    &93.89\%	  &93.87\%	 &88.14\% \\
  \hline
  5                            &91.67\%	       &85.16\%	    &95.71\%	    &95.85\%	  &95.87\%	 &93.16\% \\
  \hline
  6                            &92.76\%	       &86.19\%	    &96.96\%	    &96.62\%	  &96.52\%	 &95.92\% \\
  \hline
  7                            &93.09\%	       &86.89\%	    &97.35\%	    &97.23\%	  &97.21\%	 &96.44\% \\
  \hline
  \multirow{2}{*}{\diagbox{$k$}{states}}   & \multicolumn{3}{|c}{six-qubit (Hid-neu:1500-1500)}      &\multicolumn{3}{|c|}{seven-qubit (Hid-neu:1000-400) }     \\
  \cline{2-7}
                               &GHZ        &C23      &W          &Dicke          &$|\varphi_6\rangle$ &$|\varphi_7\rangle$ \\
  \hline
  2                            &98.28\%	   &92.24\%	 &92.91\%	 &90.46\%	     &79.17\%	          &83.44\%\\
  \hline
  3                            &99.04\%	   &93.48\%	 &94.86\%	 &96.69\%	     &86.89\%	          &89.90\%\\
  \hline
  4                            &99.42\%	   &94.69\%	 &97.8\%	 &97.82\%	     &91.52\%	          &93.47\%\\
  \hline
  5                            &99.64\%	   &95.25\%  &98.24\%	 &98.07\%	     &94.09\%             &95.04\%\\
  \hline
  6                            &99.79\%	   &97.11\%	 &98.79\%	 &98.19\%	     &95.21\%	          &96.42\% \\
  \hline
  7                            &99.87\%	   &97.41\%	 &99.22\%	 &99.01\%	     &97.10\%	          &97.16\%\\
  \hline
  \end{tabular}
  \end{center}
\end{table*}

\section{Applications of our neural networks}
How can we use a trained neural network to determine whether the fidelity of an input quantum state is higher than 96\%. The fidelity of this quantum state is first predicted using a neural network with $k=2$. If the upper bound of the predicted fidelity range given does not exceed 96\%, the neural network gives the prediction that determines that the fidelity of this state does not exceed 96\%. If the lower bound of the predicted fidelity range is more than 96\%, the neural network will give a prediction that the fidelity of this state is more than 96\%. If the range of prediction fidelity is given including 96\%, it is necessary to determine whether the prediction accuracy reaches $\pm1\%$.
If the prediction accuracy reaches $\pm1\%$, the neural network determines that the fidelity of this state does not exceed 96\%. If the prediction accuracy does not reach $\pm1\%$, the neural network cannot determine that the fidelity of this state does not exceed 96\%. Therefore, we continue the prediction with the $k=3$ neural network and repeat the above test process. Finally, the neural network can determine whether the fidelity of this state exceeds 96\%.

\section{Discussion on the scalability of the neuron number}
In the main text, we discuss the case $M=2^n-1$, and in this supplementary section we discuss the case $M<2^n-1$. When $n$ is large, $2^n-1$ will be very large, and it is impossible and unnecessary for us to compute all expectation values of the $2^n-1$ non-trivial Pauli operators and take them as neuron inputs; we can select only some of them for neuron inputs. For example, we can select those Pauli operators that contain no more than three Identity operators. There are C(n,4)+C(n,3)+C(n,2)+C(n,1)+C(n,0) such Pauli operators, so $M$ will be set to $\frac{1}{24}(n^4-2n^3+11n^2+14n)+1$. If we set $k$ to $2n$, the number of input neurons $k\times M$ becomes $\frac{1}{12}(n^5-2n^4+11n^3+14n^2)+2n$, only of the order of $n^5$ and does not increase exponentially with $n$. Figure 7 shows how the performance of our neural network fidelity estimation changes as $n$ increases from 4 to 6 for the case of $M=\frac{1}{24}(n^4-2n^3+11n^2+14n)+1$, $k=2n$ and $1-\delta=95\%$ (or 99\%). We can see that in this case the error of fidelity estimation do not increase as $n$ increases.
\begin{figure*}[ht]
  \centering
  \includegraphics[width=0.5\textwidth]{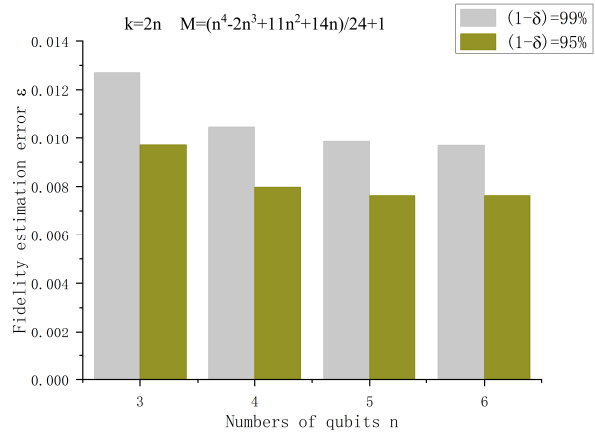}
  \caption{(Color Online) The performance of this method for fidelity estimation with neural networks varies with the number of qubits in the quantum state. The fidelity estimation error $\epsilon$ do not increase as the number of qubits $n$ increases for the case of $M=\frac{1}{24}(n^4-2n^3+11n^2+14n)+1$, $k=2n$ and $1-\delta=95\%$ (or 99\%).}
\end{figure*}


%

\begin{thebibliography}{10}
\bibitem{68three}
For 122 labels, the division of the fidelity is $[0:0.05:0.6,0.61:0.01:0.8,1-\frac{1.78}{9}:\frac{0.02}{9}:1]$.
\end{thebibliography}
\begin{thebibliography}{10}
\bibitem{67three}
For 66 labels, the division of the fidelity is $[0:0.1:0.6,0.61:0.02:0.8,1-\frac{0.98}{5}:\frac{0.02}{5}:1]$. For 122 labels, the division of the fidelity is $[0:0.05:0.6,0.61:0.01:0.8,1-\frac{1.78}{9}:\frac{0.02}{9}:1]$. For 234 labels, the division of fidelity is $[0:0.025:0.6,0.605:0.005:0.8,1-\frac{3.38}{17}:\frac{0.02}{17}:1]$. A selection of 66 labels, 122 labels and 234 labels is convenient for comparing results as the error bar can be set to $\pm 1\%$.
\bibitem{5states}
Here, we show the general states $|\varphi_2\rangle$, $|\varphi_3\rangle$, $|\varphi_4\rangle$, $|\varphi_5\rangle$, $|\varphi_6\rangle$, $|\varphi_7\rangle$ and $|\varphi_8\rangle$ as follows.
\end{thebibliography}

\begin{eqnarray*}
|\varphi_2\rangle = [0.0405;-0.4803 - 0.3933i;-0.0253 + 0.6459i;-0.3755 + 0.2327i]
\end{eqnarray*}

\begin{figure*}[hb]
\begin{eqnarray*}
\begin{array}{l}
\displaystyle
|\varphi_3\rangle = [0.1187;0.0345 + 0.0223i-0.1228 + 0.3747i;0.5924 - 0.2336i;-0.0357 + 0.4961i; \\
\displaystyle \qquad \qquad 0.1944 + 0.2406i;-0.0454 - 0.0279i;0.0690 + 0.2696i]
\end{array}
\end{eqnarray*}
\end{figure*}

\begin{figure*}[hb]
\begin{eqnarray*}
\begin{array}{l}
\displaystyle
|\varphi_4\rangle =  [0.1392;0.2107 + 0.0414i;-0.2524 - 0.2647i;0.2062 - 0.1354i;0.1718 - 0.0116i;0.1064 + 0.1494i;\\
\displaystyle \qquad\qquad 0.0789 - 0.0398i;-0.2188 + 0.3383i;-0.0520 - 0.1559i;-0.1901 + 0.1452i;-0.0245 - 0.4181i;\\
\displaystyle \qquad\qquad 0.1251 + 0.3846i;-0.1304 + 0.0025i;0.0941 + 0.0955i;-0.0577 - 0.1087i;-0.1465 + 0.1079i]
\end{array}
\end{eqnarray*}
\end{figure*}

\begin{figure*}[hb]
\begin{eqnarray*}
\begin{array}{l}
\displaystyle
|\varphi_5\rangle =[0.1592;0.0474 - 0.1834i;-0.0431 - 0.1664i;-0.3363 - 0.0787i;-0.0380 - 0.1645i;  0.0378 - 0.0606i;\\
\displaystyle \qquad\qquad -0.0846 + 0.2026i;0.1586 + 0.2416i;0.1116 - 0.0157i;0.1064 + 0.0906i;0.0608 - 0.1556i;0.0302 - 0.0418i; \\
\displaystyle \qquad\qquad -0.0254 - 0.0624i;-0.0319 - 0.2507i;0.0381 - 0.1123i;-0.1093 - 0.1652i;-0.0873 - 0.1461i;0.1807 - 0.1654i;\\
\displaystyle \qquad\qquad 0.2658 + 0.0427i;-0.2010 - 0.0322i; 0.0356 - 0.0686i;0.1222 + 0.0009i;0.0944 + 0.1331i;0.0080 + 0.0417i;\\
\displaystyle \qquad\qquad -0.0748 - 0.0483i;  0.0762 + 0.0756i;0.1541 - 0.0227i;-0.1414 + 0.1879i;-0.1427 + 0.0462i;-0.1530 - 0.0816i;\\
\displaystyle \qquad\qquad 0.0003 - 0.1342i;-0.1308 - 0.0317i]
\end{array}
\end{eqnarray*}
\end{figure*}

\begin{figure*}[hb]
\begin{eqnarray*}
\begin{array}{l}
\displaystyle
|\varphi_6\rangle =[0.1258;-0.1289 - 0.0911i;-0.0347 + 0.0199i;-0.0120 - 0.0352i;-0.0198 + 0.1522i;0.0906 + 0.0063i;\\
\displaystyle \qquad\qquad 0.0262 + 0.1211i;-0.0544 - 0.0378i;-0.0938 - 0.0502i;0.1184 - 0.0093i;0.0254 - 0.0371i;0.0822 + 0.0523i;\\
\displaystyle \qquad\qquad 0.0656 - 0.1028i;-0.0115 - 0.0604i;0.1542 + 0.1956i;0.0795 - 0.1196i;-0.0635 - 0.1297i;-0.0263 + 0.0456i;\\
\displaystyle \qquad\qquad -0.0209 + 0.1140i;-0.0203 + 0.1012i; 0.0056 - 0.1283i;0.0080 - 0.1463i;-0.0337 + 0.0195i;0.1088 + 0.0836i;\\
\displaystyle \qquad\qquad -0.0252 - 0.1061i;0.0056 + 0.0136i;-0.0605 - 0.1446i;-0.0780 - 0.0079i;0.0728 - 0.0672i;0.0632 - 0.0174i;\\
\displaystyle \qquad\qquad 0.0429 + 0.1532i;0.1260 + 0.1066i;-0.0437 - 0.2037i;0.0194 - 0.1971i;0.0806 - 0.1285i;0.0289 + 0.0105i;\\
\displaystyle \qquad\qquad 0.1379 - 0.0677i;-0.0626 + 0.0409i;-0.0276 + 0.0982i;-0.1363 - 0.1521i;0.0483 + 0.0272i;-0.1545 - 0.0268i;\\
\displaystyle \qquad\qquad -0.0652 - 0.0775i;-0.1179 + 0.1717i;0.0720 - 0.0315i;-0.0931 + 0.1785i;0.0109 + 0.1410i;0.0057 + 0.0447i;\\
\displaystyle \qquad\qquad -0.0246 + 0.0696i;-0.0725 - 0.0747i;0.0096 - 0.0414i;0.0934 - 0.0460i;-0.1239 - 0.1170i;-0.0751 + 0.1246i;\\
\displaystyle \qquad\qquad -0.1074 + 0.0287i;0.1096 + 0.1297i;-0.0941 - 0.0287i;-0.0138 - 0.0076i;-0.0870 + 0.1829i;-0.0824 + 0.0722i;\\
\displaystyle \qquad\qquad 0.1162 + 0.0492i;0.0326 + 0.0584i;-0.0073 + 0.0186i;-0.0626 - 0.0719i]
\end{array}
\end{eqnarray*}
\end{figure*}

\begin{figure*}[hb]
\begin{eqnarray*}
\begin{array}{l}
\displaystyle
|\varphi_7\rangle =[0.0695 + 0.0000i;0.0605 + 0.0281i;0.0065 - 0.0543i;-0.0608 - 0.0294i;-0.0193 - 0.0656i;0.0126 - 0.0140i;-0.0606 + 0.0615i;\\
\displaystyle \qquad -0.0246 - 0.0569i;-0.0276 + 0.0123i;0.0251 - 0.0021i;-0.0825 - 0.0712i;-0.0598 + 0.0683i;0.0365 + 0.1367i;0.0548 - 0.0188i;\\
\displaystyle \qquad -0.0003 - 0.0099i;0.0557 - 0.0037i;0.0968 + 0.0779i;-0.0899 + 0.0037i;-0.0453 - 0.1019i;0.0547 - 0.0415i;0.0187 - 0.0574i;\\
\displaystyle \qquad -0.0894 - 0.0183i;0.0345 + 0.1262i;-0.0366 + 0.0623i;0.0812 - 0.0488i;-0.0056 - 0.0852i;-0.0227 - 0.0584i;0.0287 + 0.1105i;\\
\displaystyle \qquad  0.0044 + 0.0009i;-0.0175 + 0.1007i;-0.0397 - 0.0185i;-0.0598 - 0.0394i;0.0136 - 0.0341i;0.0892 + 0.0635i;0.0696 + 0.0376i;\\
\displaystyle \qquad 0.0592 - 0.0695i;0.0122 - 0.0334i;0.0016 + 0.0295i;0.1484 + 0.0320i;0.0517 - 0.0903i;0.0249 + 0.0312i;-0.0375 - 0.1190i;\\
\displaystyle \qquad -0.0739 - 0.0196i;-0.0313 - 0.0611i;-0.0138 + 0.0379i;0.0810 - 0.0439i;-0.0471 - 0.0356i;-0.0764 - 0.0836i;-0.0162 - 0.0124i;\\
\displaystyle \qquad 0.0100 + 0.0064i;0.0033 + 0.0585i;0.1609 + 0.0219i;0.0042 + 0.0286i;-0.0129 + 0.0069i;0.0504 - 0.0814i;0.0370 - 0.0630i;\\
\displaystyle \qquad  0.0743 + 0.1463i;-0.0684 + 0.0164i;-0.0886 - 0.0349i;0.0061 - 0.0597i;-0.0324 - 0.1242i;-0.0091 + 0.0008i;0.0787 + 0.0286i;\\
\displaystyle \qquad  0.0409 - 0.0340i;-0.0239 + 0.1039i;-0.0573 - 0.0595i;0.0792 + 0.0325i;0.0242 - 0.0019i;0.0699 + 0.0603i;-0.0480 - 0.0524i;\\
\displaystyle \qquad -0.0488 + 0.0837i;-0.0117 + 0.0479i;-0.0547 + 0.1122i;-0.0057 - 0.0231i;0.0613 + 0.0400i;-0.1400 + 0.0209i;0.0106 + 0.0753i;\\
\displaystyle \qquad  0.0334 - 0.0111i;-0.0534 - 0.0892i;0.0054 - 0.0649i;0.0865 + 0.0146i;-0.0469 + 0.0326i;0.1115 - 0.0396i;0.0638 - 0.1242i;\\
\displaystyle \qquad  -0.0197 + 0.0983i;0.0072 - 0.0617i;-0.0587 - 0.0427i;-0.0126 + 0.0883i;-0.0648 + 0.0306i;0.0013 + 0.0208i;-0.1021 + 0.0107i;\\
\displaystyle \qquad  -0.0636 - 0.0758i;0.0345 + 0.0477i;0.0010 + 0.0965i;-0.0204 - 0.0327i;0.0474 - 0.0233i;0.1390 - 0.0374i;0.0054 + 0.1080i;\\
\displaystyle \qquad  -0.0904 + 0.0029i;-0.1118 + 0.0338i;0.0609 - 0.0188i;-0.0358 + 0.0012i;0.0598 + 0.0508i;-0.0123 + 0.0006i;0.0114 - 0.0653i;\\
\displaystyle \qquad  -0.0517 + 0.0835i;0.0115 - 0.0751i;-0.0028 + 0.0802i;0.0119 + 0.0615i;0.0752 + 0.0779i;0.0321 + 0.0072i;0.0534 - 0.0179i;\\
\displaystyle \qquad   0.0717 + 0.0393i;0.0414 + 0.0350i;0.0897 + 0.0488i;0.1269 + 0.0272i;0.0059 - 0.0434i;-0.1644 - 0.0756i;0.0114 + 0.1916i;\\
\displaystyle \qquad   0.0348 + 0.0125i;0.0689 + 0.0668i;0.0453 - 0.0889i;0.0773 + 0.0467i;-0.0038 - 0.0707i;0.0586 - 0.0205i;-0.0463 + 0.0566i;\\
\displaystyle \qquad   0.1528 + 0.0668i;0.1423 - 0.1300i]
\end{array}
\end{eqnarray*}
\end{figure*}

\begin{figure*}[hb]
\begin{eqnarray*}
\begin{array}{l}
\displaystyle |\varphi_8\rangle =[0.0608241482079368 + 0.00000000000000i;0.118916150247069 - 0.0334017340293308i;\\
\displaystyle \qquad0.0613508450188904 + 0.000699684679667262i;0.0377942962466794 + 0.00760773380239735i;\\
\displaystyle \qquad0.0518554511638503 - 0.0629347710689567i;-0.0348903567710464 - 0.0718549996379180i;\\
\displaystyle \qquad0.0227098419086832 - 0.00154420922029817i;-0.0497191518722269 - 0.0243340000060727i;\\
\displaystyle \qquad0.0191052748480883 + 0.0127099467287183i;0.0815322723601795 - 0.0173207223850301i;\\
\displaystyle \qquad-0.0386203561726839 + 0.0393996047287102i;-0.0843992334500119 + 0.0262117932551506i;\\
\displaystyle \qquad-0.0205797695445411 - 0.0408110927023867i;-0.0719368273356829 + 0.0289919253060471i;\\
\displaystyle \qquad-0.00509072143998464 + 0.0187080212356307i;0.0156572885903232 + 0.0587429135229369i;\\
\displaystyle \qquad-0.0789073759826587 + 0.0434789869561237i;-0.0479199964481745 + 0.0201699941139512i;\\
\displaystyle \qquad0.00990911591493497 + 0.00903046321813193i;-0.108281939042704 + 0.0262738826617960i;\\
\displaystyle \qquad0.0286201532924058 - 0.0321510933788634i;-0.00594281098237468 - 0.0277837309503670i;\\
\displaystyle \qquad-0.00897263434177392 - 0.00286188953397132i;-0.0475376700270029 - 0.0534302182935278i;\\
\displaystyle \qquad-0.0756403239274263 + 0.0476354532957176i;-0.107509142858203 - 0.0610222801571933i;\\
\displaystyle \qquad-0.0421017454667555 + 0.0248590755850056i;0.0330235733307558 + 0.0263171656622000i;\\
\displaystyle \qquad-0.00115134907900766 - 0.0569471023082743i;-0.0600399027302497 - 0.0601866941036766i;\\
\displaystyle \qquad-0.0581951627223363 - 0.0966804757330855i;0.0298763370172347 - 0.0369286927860094i;\\
\displaystyle \qquad0.0547427659586129 - 0.0267006115448212i;0.0404026859581070 - 0.0176862364549503i;\\
\displaystyle \qquad-0.0540554719377302 - 0.0490586969743824i;-0.00567232966829234 - 0.0816034820597759i;\\
\displaystyle \qquad-0.0320472389148351 - 0.0276355677495632i;0.0298823925247719 + 0.00805073438496761i;\\
\displaystyle \qquad-0.0692942681051767 - 0.0321306598059461i;0.0113271228127193 + 0.00627492186794386i;\\
\displaystyle \qquad-0.00562930824881366 + 0.0197578122305785i;0.0685312688977480 + 0.0450049223530517i;\\
\displaystyle \qquad-0.0331418541087130 - 0.0701520700436068i;0.0175639066789611 + 0.0250082533594779i;\\
\displaystyle \qquad0.00371395016948587 + 0.0290571452353454i;0.0275747221135126 - 0.0114820569920976i;\\
\displaystyle \qquad0.0267993392264768 + 0.136974455936479i;-0.0243418619295571 - 0.00802664379554786i;\\
\displaystyle \qquad0.0561362781129846 + 0.0211851142226193i;0.0751527285796857 - 0.0513162905397350i;\\
\displaystyle \qquad0.00649683053731943 + 0.0142036959661945i;-0.0740464534612519 - 0.0125076812872711i;\\
\displaystyle \qquad-0.0270832641582373 - 0.0661334409088159i;-0.0609969009081254 + 0.0120628402025501i;\\
\displaystyle \qquad0.0645873237436956 + 0.0351469725680645i;-0.0126566148581343 - 0.0530495258617872i;\\
\displaystyle \qquad0.0444113060675021 - 0.00302815938772813i;-0.0443889412931621 - 0.0311443629121288i;\\
\displaystyle \qquad0.0116317857628378 - 0.0777738725355282i;0.0396866189033302 - 0.00609411105956391i;\\
\displaystyle \qquad-0.0195451763601245 + 0.0204381872763298i;-0.0198966131778964 - 0.00629058894740721i;\\
\displaystyle \qquad-0.0521935121563282 - 0.0182601955075468i;0.0594020408300149 - 0.00924562646445881i;\\
\displaystyle \qquad0.0126793572191845 - 0.0284762224113334i;-0.0450072887070507 - 0.00119129789017963i;\\
\displaystyle \qquad-0.0143420133612828 + 0.0498139168126091i;-0.0655440036983153 - 0.00669563514666392i;\\
\displaystyle \qquad-0.0309318934263726 - 0.00156931932103170i;-0.0332184485474387 - 0.0383272076954024i;\\
\displaystyle \qquad0.0837167873960070 + 0.0463523697650752i;0.0176889334942217 + 0.0108377990950843i;\\
\displaystyle \qquad0.0330990695030220 + 0.0189503490163052i;0.100938438807977 - 0.103972241184801i;\\
\displaystyle \qquad-0.0383588637239730 + 0.0406733193256555i;-0.0323442360058115 - 0.00289573819601914i;\\
\displaystyle \qquad0.00223770822472600 - 0.00824976606431206i;-0.0580370416971857 - 0.000735636659276832i;\\
\displaystyle \qquad-0.0165303006364171 - 0.0188466885897068i;0.0450493601343480 - 0.0280401506721841i;\\
\displaystyle \qquad-0.0236578399413773 + 0.0392584607930664i;-0.0331847494509494 - 0.0119676717087829i;\\
\displaystyle \qquad0.0295931064018567 + 0.0305065048945616i;-0.0152255287111828 - 0.00485029917598210i;\\
\displaystyle \qquad0.0243377609939934 + 0.0496753699283269i;0.0104220576600195 - 0.0699835433562266i;\\
\displaystyle \qquad0.0309522464926962 - 0.0478670215935720i;-0.00710864428077468 + 0.000862329175744420i;\\
\displaystyle \qquad0.0389774368641183 + 0.0165826116909666i;-0.0536370774304833 - 0.000513989621822980i;\\
\displaystyle \qquad-0.00166482528680707 - 0.0255077253174407i;-0.000299021074767773 + 0.0280597818031829i;\\
\displaystyle \qquad-0.0156037776886095 + 0.0486233488316806i;0.0295627687846105 - 0.0190813140657321i;\\
\displaystyle \qquad-0.0394074641702558 - 0.0438108225966595i;-0.000710014204684279 - 0.0156113130716522i;\\
\displaystyle \qquad0.0237497062145224 - 0.00743196587411630i;-0.0425565788785281 - 0.0300639209878852i;\\
\displaystyle \qquad-0.0390249131567994 + 0.0683279573191321i;-0.0651570033275084 + 0.0413915756041366i;\\
\displaystyle \qquad0.0418211998636496 + 0.0528452859827522i;-0.00277330209045970 + 0.0408031605450579i;\\
\displaystyle \qquad-0.0890891775259274 + 0.0618620810395485i;-0.127544861675430 - 0.0164855842327497i;\\
\displaystyle \qquad0.110600759493332 + 0.0795666115662246i;-0.0906072327543428 - 0.0125888879644459i;\\
\displaystyle \qquad0.0482137187533687 + 0.0102106002652661i;0.0238197102537148 - 0.0746563424459331i;\\
\displaystyle \qquad-0.00518561593302128 - 0.00838737840451845i;-0.0150765111385722 - 0.0141903598665584i;\\
\displaystyle \qquad0.0264755373005631 - 0.0366053990165149i;0.0525812314523336 - 0.00934062967673553i;\\
\end{array}
\end{eqnarray*}
\end{figure*}
\begin{figure*}[hb]
\begin{eqnarray*}
\begin{array}{l}
\displaystyle \qquad0.0713214631130962 + 0.0409922490239104i;-0.0200505747861032 - 0.0998875439569777i;\\
\displaystyle \qquad-0.0890740085951002 - 0.0451440605434097i;0.0220803106440334 + 0.00904119775960116i;\\
\displaystyle \qquad0.0372215288058124 + 0.0231620988354949i;-0.0505881780660252 + 0.0332152882051352i;\\
\displaystyle \qquad0.0740115678486559 + 0.0134403902077525i;-0.0402477175162510 - 0.0333961942486656i;\\
\displaystyle \qquad-0.0165353328495778 - 0.00552536182922306i;-0.0372677398038348 - 0.00213191952001035i;\\
\displaystyle \qquad-0.0433376192983931 + 0.0198976504573799i;0.0384114459041732 + 0.106279330949868i;\\
\displaystyle \qquad0.0459274112868497 - 0.0162370774141051i;0.0676312205493608 + 0.0537714026643900i;\\
\displaystyle \qquad0.00352196444577846 + 0.0329091153400323i;-0.0207326779510939 + 0.0163956536771018i;\\
\displaystyle \qquad0.0218991061748520 - 0.0940646040108595i;-0.0536124232943482 - 0.00934249394989665i;\\
\displaystyle \qquad-0.0398381591273541 + 0.0116824289432429i;-0.0323504031970830 + 0.00388858340951971i;\\
\displaystyle \qquad-0.00456059197280053 + 0.0364889637016544i;-0.0261495741815767 + 0.0169111815624359i;\\
\displaystyle \qquad0.0111165843576871 - 0.0247044337096012i;-0.00264515149284518 - 0.0470163563489267i;\\
\displaystyle \qquad-0.0417150816276546 + 0.0313923456286448i;0.0514064311352219 + 0.00141579058581644i;\\
\displaystyle \qquad-0.0187067873176759 + 0.0406458950225942i;0.0136558603643515 + 0.00593034688456360i;\\
\displaystyle \qquad-0.0319180514111241 + 0.0643207888708036i;0.0772127830738079 + 0.0786893188300254i;\\
\displaystyle \qquad-0.0163174333402343 + 0.0455043918993442i;0.0273326058646386 + 0.0724127677393314i;\\
\displaystyle \qquad-0.0317357579098575 + 0.0441672425931433i;-0.0146158933058286 + 0.0889401520236187i;\\
\displaystyle \qquad-0.0583998642017608 + 0.0783183829521906i;-0.00576901939085139 - 0.00137955197989608i;\\
\displaystyle \qquad0.0692818479279746 - 0.0175350456506586i;-0.0642796861414790 - 0.0642251459700653i;\\
\displaystyle \qquad0.0368220469004493 + 0.0624548268795734i;0.0420145561921296 - 0.0328403709249718i;\\
\displaystyle \qquad0.00123595123345024 + 0.0325910340982914i;-6.02777585948601e-05 - 0.0362123605535817i;\\
\displaystyle \qquad-0.0732796776049777 - 0.0308817049879325i;-0.00948713265323484 - 0.0796343873815945i;\\
\displaystyle \qquad0.0443263142034812 - 0.0578734796623575i;0.0118969746827515 - 0.0150484998853584i;\\
\displaystyle \qquad0.0338160181701561 + 0.0748360333598817i;-0.0739924799764414 + 0.0363690446484240i;\\
\displaystyle \qquad0.0159901277852531 - 0.0154779763340079i;-0.0446109734793101 + 0.00644608512767663i;\\
\displaystyle \qquad0.0134237115470127 - 0.0381284179286567i;0.0872875509837861 - 0.0233217837313136i;\\
\displaystyle \qquad0.0127140346721947 + 0.0215622465818749i;0.0227155413916561 - 0.0634255277231416i;\\
\displaystyle \qquad-0.0157939141814109 + 0.0876881305175142i;-0.117125186512107 - 0.0564758364542242i;\\
\displaystyle \qquad0.00874557533942336 + 0.0431209699792748i;-0.0111721688598814 - 0.0327436440461095i;\\
\displaystyle \qquad-0.0181770357560778 + 0.0189505789204447i;0.0411790437777966 - 0.0791320244039931i;\\
\displaystyle \qquad0.0482390834042627 - 0.0116143531219595i;-0.0360281987407459 + 0.0400059961374163i;\\
\displaystyle \qquad-0.0201672444019088 - 0.0110101400241323i;0.0497454785664465 + 0.0445114040098613i;\\
\displaystyle \qquad-0.0157625288571732 + 0.0417675031009961i;-0.0295497289161719 + 0.0493213071930167i;\\
\displaystyle \qquad-0.0473207230398064 - 0.0204185147464490i;-0.0370871648327222 - 0.00303045857063023i;\\
\displaystyle \qquad-0.00872784729771407 + 0.00772167217671842i;0.0538689068502806 - 0.0264283743764751i;\\
\displaystyle \qquad0.0139435772843981 - 0.0189107733133991i;-0.0107476944585334 + 0.0340851083448936i;\\
\displaystyle \qquad-0.00589438133141103 - 0.0715190643921810i;0.0301192219750676 + 0.0517376981308512i;\\
\displaystyle \qquad0.00751670147131148 + 0.0409022496334675i;-0.0794063974447763 + 0.0336612017478651i;\\
\displaystyle \qquad0.117146899476949 + 0.0239461610824425i;0.0536074011518861 - 0.0267049703797047i;\\
\displaystyle \qquad0.0197194606265434 - 0.00831707640288587i;-6.88810640741180e-05 + 0.00606402612988709i;\\
\displaystyle \qquad0.0169455920444588 - 0.0520887355242409i;-0.0503875576198242 + 0.0650709969181780i;\\
\displaystyle \qquad-0.0159168111032638 - 0.0260907762866937i;-0.0599573573995700 + 0.0597072323554902i;\\
\displaystyle \qquad-0.0553062988300955 + 0.0309190206617516i;0.0157176950898955 - 0.0174880409124464i;\\
\displaystyle \qquad0.0226419337858581 + 0.0312767025057463i;-0.0309202224479869 + 0.00271792003791919i;\\
\displaystyle \qquad-0.0146101619808344 - 0.0480313734544479i;0.0312894855955269 - 0.0222182293575639i;\\
\displaystyle \qquad0.0488853696936001 + 0.0520533273519843i;0.0329068647852443 - 0.0697680181239380i;\\
\displaystyle \qquad-0.0348029758961616 + 0.0189247755179659i;0.0529445591977682 + 0.0244805018238514i;\\
\displaystyle \qquad-0.0303162165278065 + 0.0941231333202816i;0.0349945828625536 + 0.0130680700752814i;\\
\displaystyle \qquad-0.0193885008966607 + 0.0621662487362271i;0.0540248131475878 - 0.0202303522689034i;\\
\displaystyle \qquad0.0310175265290909 + 0.0554971618687696i;0.0101773423187402 - 0.000810457427479659i;\\
\displaystyle \qquad-0.0143284334874021 + 0.0132193279956417i;0.0334971783870023 + 0.0119762543103859i;\\
\displaystyle \qquad-0.0705783103058466 + 0.0391556694037088i;-0.0204689712663568 + 0.0291419630657466i;\\
\displaystyle \qquad-0.00997272321664950 - 0.0137058274534058i;0.00436636920955455 - 0.0299336320762942i;\\
\displaystyle \qquad-0.00468001642082807 - 0.000130839654754422i;-0.00254453687001702 - 0.000469824785165419i;\\
\displaystyle \qquad0.00721764988760490 - 0.0702294325510927i;0.0317390462166900 - 0.100853280710839i;\\
\displaystyle \qquad-0.00563863667827467 - 0.0486207777031203i;0.0466194926982029 - 0.0203464021645670i;\\
\displaystyle \qquad0.0184878333169885 - 0.00418919844322236i;-0.0461700143077590 + 0.0402148662740331i;\\
\displaystyle \qquad-0.0197737003579020 + 0.00158175460684209i;0.0285725233769325 - 0.0212608171445315i;\\
\displaystyle \qquad0.104047625981961 - 0.0454235856022888i;-0.0636522416142955 + 0.0243775870661680i;\\
\end{array}
\end{eqnarray*}
\end{figure*}
\begin{figure*}[hb]
\begin{eqnarray*}
\begin{array}{l}
\displaystyle \qquad-0.0101648768928627 + 0.0331897924925792i;0.00670799944140863 - 0.0297970810625649i;\\
\displaystyle \qquad-0.00246131181036723 + 0.123835563661943i;0.0483799680508918 + 0.0275892116505744i;\\
\displaystyle \qquad-0.0581480003790039 - 0.000440997264230780i;-0.0970482629830606 - 0.000451618587710918i;\\
\displaystyle \qquad0.0380632546108555 + 0.0164558771525070i;0.0134593662715805 + 0.101741425663603i;\\
\displaystyle \qquad-0.00659308867087655 - 0.0629865434895194i;-0.102338537537638 - 0.00352999686667835i;\\
\displaystyle \qquad-0.0700726754997977 + 0.0449751536414556i;0.0560565609209647 - 0.000777283791904981i;\\
\displaystyle \qquad0.00372900569268532 - 0.0582681127011097i;0.0344340234372171 + 0.0389547026670941i;\\
\displaystyle \qquad0.0719595444205174 + 0.0259927978298189i;0.0375303951255054 + 0.0563237809418214i;\\
\displaystyle \qquad0.0249933409592011 + 0.0528500773965271i;-0.0511460957023439 + 0.00776994916626157i;\\
\displaystyle \qquad-0.00845594085081976 - 0.0156293278113676i;0.00936729435494009 + 0.0672539846043160i;\\
\displaystyle \qquad-0.0366299250819762 + 0.00223760377252909i;-0.0225389293250788 + 0.0502738786426404i;\\
\displaystyle \qquad0.00844308104055903 - 0.0232542239805346i;-0.0106150709900326 + 0.00234139813519379i;\\
\displaystyle \qquad 0.0602206456293255 - 0.0100848833161413i;-0.00867374519320297 - 0.0544267930770106i]
\end{array}
\end{eqnarray*}
\end{figure*}

\end{document}